%% file: main.tex
\documentclass[11pt]{article}

\usepackage{graphicx}
\usepackage{caption,subcaption}
\usepackage{gnuplot-lua-tikz}

\usepackage[colorlinks=true,linkcolor=red,urlcolor=blue]{hyperref}

\usepackage{cite}

\usepackage{setspace}
\onehalfspace

\begin{document}
\begin{center}
{\huge Measurement sensitivity and resolution\\for background oriented schlieren during image recording\footnote{The original manuscript is accepted for publication by Journal of Visualization, and it is available at www.springerlink.com.}}

\vspace{16pt}
{\sc Ardian B. Gojani\footnote{Institute of Fluid Science, Tohoku University, gojani@edge.ifs.tohoku.ac.jp. Author acknowledges the support of GCOE program at Tohoku University.} \& Burim Kamishi\footnote{Department of Physics, Faculty of Mathematical and Natural Sciences, University of Prishtina.} \& Shigeru Obayashi\footnote{Institute of Fluid Science, Tohoku University.}}

\vspace{16pt}
\end{center}

\begin{spacing}{1}
\noindent \textbf{Abstract:} Background oriented schlieren (BOS) visualization technique is examined by means of optical geometry. Two most important results are the calculation of the sensitivity and spatial resolution of a BOS system, which allows for the determination of the experiment design space. A set of relations that characterize the performance of a BOS measurement is given, with emphasis on the design of background pattern and spatial placement of optical components.
\end{spacing}

\section{Introduction}
\label{intro}
Background oriented schlieren (BOS) is a flow visualization technique that carries the promise of quantitative measurements, as introduced by Meier (2002) and Richard and Raffel (2001). The physical principle on which BOS relies is similar to that of the formation of mirages: a scene viewed through a hot plume of air will appear distorted. The distortion is caused by the bending of the light rays from the straight line propagation due to inhomogeneities in the medium, and it is described by the functional dependence of the refractive index on space, $n(\mathbf{r})$ (Hecht 2002). One can consider BOS as an improvement and/or simplification of defocused grid schlieren (Vasil'ev 1971), white light speckle photography (Giglio et al 1980), moir\'{e} deflectometry (Merzkirch 2007), and similar techniques. A large group of researchers refer to BOS as synthetic schlieren, with the main difference between the two being that BOS is more general in terms of the background, while synthetic schlieren employs an optimized synthetically generated pattern as a background (Dalziel et al 1998). There are two fundamental differences between BOS and standard (Toepler) schlieren techniques: in the latter, one records the variation of illumination in the image plane, and also one intervenes on the light path, usually with a knife-edge, modifying it in addition to the modifications by the flow. 

In BOS, a pattern is recorded at two different instances, once without and once with the flow that is to be investigated, giving the reference and the measurement images, respectively. The flow causes inhomogeneities in the medium, which will influence the refractive index. Thus, when the reference image is recorded, the refractive index of the medium is considered to be constant. On the other hand, when the the measurement image is recorded, the refractive index is a function of space in the domain defined as the test section. Interrelationship between the refractive index and flow properties -- mainly density, as expressed by the Lorentz-Lorenz equation -- allows for determination of the state of the fluid.\footnote{Lorentz-Lorenz equation relates polarizability of an assembly of molecules to the refractive index. For gases, this relation can be approximated by the Gladstone-Dale equation (Born and Wolf 2005).} Based on this, BOS has found many applications, as described in papers by Venkatakrishnan and Meier (2004), Kinder et al (2007), Moisy et al (2009), Mizukaki (2010), Kirmse et al (2011), Sourgen et al (2012), Glazyrin et al (2012), and Vinnichenko et al (2012), to name a few.

BOS technique consists of two independent stages: image recording and image evaluation. The first deals with the problems of designing the pattern that is to be imaged, placement of the components in the optical setup, and determining appropriate conditions for recording the images. The image evaluation part deals with the problems of comparing reference and measurement images, and data extractions from those comparisons. The dominant method for data extraction in BOS relies on cross-correlation algorithms that have been developed for speckle photography and particle image velocimetry. Alternative methods include optical flow (Atcheson et al 2009), laser interferometric computed tomography (Ota et al 2011), and recently some developments on single pixel correlation are being proposed (K\"{a}hler et al 2012). 

There have been several publications, including those by Elsinga et al (2004), Goldhahn and Seume (2007), Yevtikhiyeva et al (2009), Ambrosini et al (2012), and Gojani and Obayashi (2012), which assessed the sensitivity and accuracy of BOS technique, mainly from the data processing and image evaluation point of view, because these introduce the dominant measurement uncertainties. Nevertheless, in order to determine the performance of a BOS system, it is necessary to investigate the measurement uncertainties arising from the optical setup, as well. This becomes more important in the cases when experimentation is carried with instruments of modest specifications, such as the case of high-speed imaging, limited by the relatively low pixel count of the image sensor.

\section{Sensitivity}
\label{sens}
\begin{figure}
\input{fig-bossetup.tex}
\caption{Optical setup for background oriented schlieren.}
\label{fig:bossetup}
\end{figure}
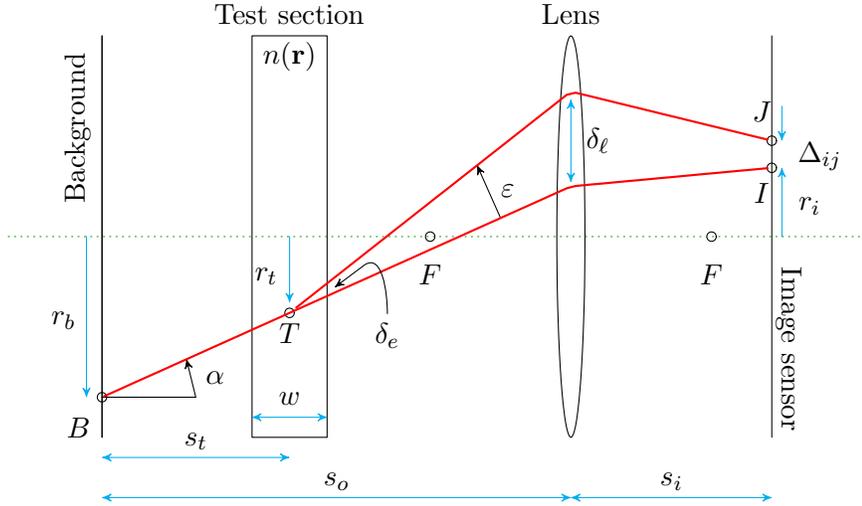

BOS measurement is able to quantify ray deviation, but only integrated through the line of sight. This is illustrated in figure \ref{fig:bossetup}, which shows the meridional plane of a typical BOS setup consisting of a structured background, the test section under investigation, (also referred to as the phase object, transfer channel, density field, etc.), the objective lens with focal length $f$ (focal points $F$), the image recording sensor, and the respective distances. When the reference image is being recorded, a feature from the background located at the point $B$ is imaged in the point $I$. Introduction of the fluid flow with variable refractive index $n(\textbf{r})$ will deflect the beam for an angle $\varepsilon$, thus the imaged point now will be shifted for $\Delta_{ij}$ to the point $J$. A necessary condition for this effect to be observed is that the pattern shift $\Delta_{ij}$ be larger than the pixel linear dimension $\ell_{px}$. Thus, the detection limit is determined based on the relation\footnote{This equations holds for a ray of light that images a point from the physical space into a point in image space. As it is known from image analysis, the determination of pixel shifts with subpixel resolution is possible, but only over an area.}

\begin{equation}\label{eq:detection}
\Delta_{ij} \geq \ell_{px}.
\end{equation}

A consequence of the change of the ray direction is that the ray will exit the test section at a different point, as compared to the ray during the recording of the reference image. But, this difference, namely $\delta_e$ in figure \ref{fig:bossetup}, is negligible for as long as $\delta_{\ell}$, which is the distance between the entry points of these rays in the lens, is much larger. Since $\delta_e/ \delta_{\ell} \approx w/ (s_o-s_t)$, this condition is satisfied if
\begin{equation}\label{eq:negl}
\frac{w}{s_o-s_t} << 1.
\end{equation}
This means that in the case of a BOS setup with dimensions much larger than the width of the test section, one can assume that the fluid flow only deflects the light ray, but does not displace it. Furthermore, similarly to the thin lens which is described by the principal plane, the entire test section can be approximated by a refractive plane.

The sensitivity $S$ of a measurement setup is defined as the ratio of the detected change - for the case at hand, the pattern shift $\Delta_{ij}$, - to the corresponding change in the parameter being measured, - angle of deflection $\varepsilon$, - that is,
\begin{equation}\label{eq:sensitivity}
S = \frac{\Delta_{ij}}{\varepsilon}.
\end{equation}

Let the distance of a background point $B$ from the optical axis be $r_b$ (in the object space), as shown in figure \ref{fig:bossetup}. The lens with diameter of aperture $A$ forms the image of $B$ at point $I$ by collecting only rays emitted within the range of angles $\alpha$, for which
\begin{equation}\label{eq:angles}
\frac{r_b-A/2}{s_o} \leq \tan \alpha \leq \frac{r_b+A/2}{s_o},
\end{equation}
where $\alpha$ is the angle of the direction of the ray with respect to the optical axis, and $s_o$ is the distance between the principal plane (the plane of lens and aperture) and the background. In practice, $A \approx 1 - 50 \times 10^{-3}$ m and $s_o \approx 0.1 - 100$ m.\footnote{There are examples of using BOS technique for visualizing open air blast waves with $s_o \approx 400$ m and $s_t \approx 100$ m (Mizukaki et al 2012).} For the paraxial approximation to be applicable within an accuracy of 1\% ($\alpha < 15^{\circ}\approx 0.25$ rad), the relationship between $r_b$ and $s_o$ is 
\begin{equation}\label{eq:field}
r_b < 0.25 \, s_o,
\end{equation}
which means that the dimensions of the field of view should not exceed the quarter of the dimensions of the BOS setup.

In BOS, the thin lens relation $1/f = 1/s_i + 1/s_o$ holds, with the lens being focused on the background. The image $I$ is at a distance $r_i = M\, r_b$ from the optical axis (in the image space), where $M = s_i/s_o$ is the optical magnification of the system. Here, since BOS is basically a photographic technique of real and inverted images, the magnification is taken to be positive, as is customary in photography. In the case of the deflected ray, the image point $J$ is at a distance $r_j = M [r_t-(\alpha+\varepsilon)s_t] = r_i - M\, \varepsilon\, s_t$, where $r_t = r_b + \alpha\, s_t$ is the distance of the deflecting point $T$ from the optical axis, $s_t$ is the distance between the background and the test section (see figure \ref{fig:bossetup}).

Since $\Delta_{ij} = |r_i-r_j| = \varepsilon\, M\, s_t$, the sensitivity can be expressed as 
\begin{equation}\label{eq:sens}
S = M \, s_t = \frac{s_i \, s_t}{s_o}.
\end{equation}

\section{Spatial resolution}
\label{resol}

In BOS measurements, reference and measurement images are taken under the same lightning conditions of the field of view, which consists of a pattern with highly varying brightness from point to point. Nevertheless, irrespective of the radiant exitance from a background point, it is the irradiance on the pixel area - resulting in the image of the said point - that is actually used for image evaluation and quantitative measurements. Thus, for a linear proportionality between exitance and irradiance, it is the relative difference in irradiance from point to point in the image space that determines the pattern.

BOS imaging is done with charge-coupled device (CCD) or complementary metal-oxide semiconductor (CMOS) image sensors, which digitize the output of the light falling onto the area of a pixel. Digitalization of the image of the field of view leads to discretization in two domains: (i) in the space domain, an operation referred to as sampling, and (ii) in the domain of intensity values, an operation referred to as quantization. Quantization is influenced by several noise sources in the image sensor, but the following discussion will assume that the variation of irradiance due to imaged pattern is much larger than the combined noise levels (dark frame and individual pixel response). In a further simplification, it is assumed that the image is of a binary type; hence, when the (relatively) high irradiation covers more than half of the pixel, the pixel records light (pixel grayscale value 1), otherwise it does not (pixel grayscale value 0).

\begin{figure}
\input{fig-circleofconf2.tex}
\caption{Cone of light creating the image in BOS.}
\label{fig:circle}
\end{figure}
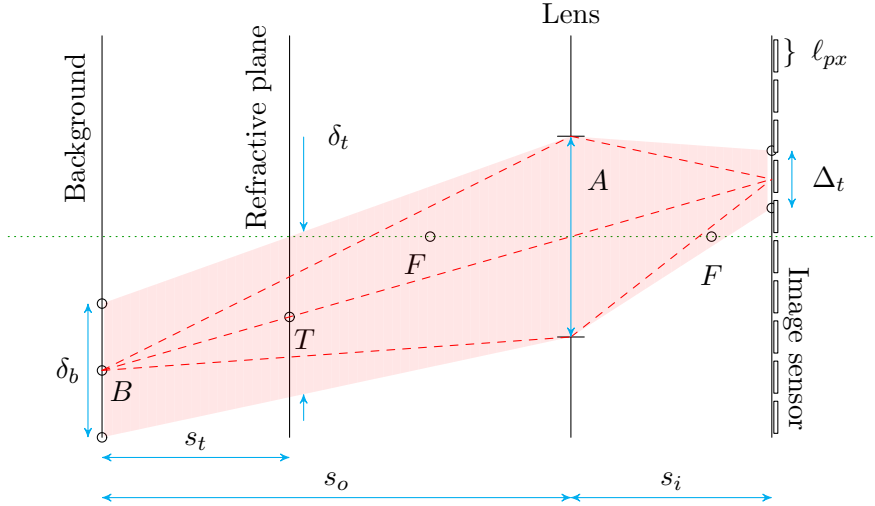

Similarly to the previous section, let the image of the background point $B$, i.e. point $I$, be at a distance $r_i$ from the optical axis, falling on the $i-th$ pixel ($i = 1, 2, \ldots, N$, where $N$ is the number of pixels of the sensor in one direction). Because of sampling, all the other points that fall in the segment
\begin{equation}\label{eq:segment}
\left(i-\frac{1}{2}\right)\ell_{px} \leq r_i \leq \left(i+\frac{1}{2}\right)\ell_{px}
\end{equation}
are indistinguishable (for analysis in 2D, the factor multiplying the length of a pixel should be $i\pm1/\sqrt{2}$, but this correction is insignificant in our case). Furthermore, let $\Delta_t = k\, \ell_{px}$ be the diameter of the circle of confusion (CoC), with $k$ a small integer (let us say, $k = 1, 2, \ldots, N/100$ pixels). Based on the properties of CoC, only irradiance variations that are at a distance larger than this diameter can be resolved, therefore all points within the segment $r_i \pm \Delta_t/2$ are indistinguishable. Then, condition (\ref{eq:segment}) is modified to
\begin{equation}\label{eq:coc}
\left\lfloor i-\frac{k+1}{2}\right\rceil \leq \frac{r_i}{\ell_{px}} \leq \left\lfloor i+\frac{k+1}{2}\right\rceil,
\end{equation}
where $\lfloor x \rceil$ designates the nearest integer function of $x$.

The cone of light collected by the aperture with diameter $A$, as it is illustrated by the shaded area in figure \ref{fig:circle}, does not contain independent information along its cross-section, because image sensor records the average irradiance contained in it, thus it appears uniform. Backprojection of the segment defined by equation (\ref{eq:coc}) through the lens into the background plane determines the smallest distance that can be resolved, $\delta_b$, which has the value 
\begin{equation}\label{eq:delta_b}
\delta_b = \frac{k+1}{M} \frac{\ell_{px}}{2},
\end{equation}
and this defines the diameter of a dot in the background, i.e. the shortest distance between pattern variations that are resolved. In practice, the dot is made two or three times larger, so that image evaluation by cross-correlation can yield an optimized result.

The spatial resolution of a BOS system at the deflection point $T$ corresponds to the diameter of the projection of the cone of light on the refractive plane, and its value is
\begin{equation} \label{eq:delta_t}
\delta_t=\frac{M}{M+1}\frac{s_t}{f_{\#}} + \frac{\Delta_t}{M}\left(1-\frac{s_t}{s_o}\right),
\end{equation}
where $f_{\#} = f/A$ is the focal ratio (f-number) of the lens, and where $s_t/s_o < 1$ holds.

\section{Discussion}
\label{disc}
In BOS, the test section is placed between the background and the objective lens. This limits the values of $s_t$ in the range $0 < s_t <s_o$, and consequently the sensitivity of the BOS setup is 
\begin{equation}
0 < S < s_i.
\end{equation}
This results shows that the sensitivity of a BOS setup can be adjusted to any desired level, limited only by the focal length of the lens. In practice, though, sensitivity is adjusted by deciding on the spatial extent of the experimental setup (adjustment of $s_t$, limited by the laboratory space and the field of view) and the lenses and camera used (adjustment of $M$). Older publications that treated the sensitivity of BOS argued for the use of long focal length lenses for increased sensitivity. While this is correct, the present paper gives a more fundamental reason on why long focal length lenses are preferred: longer focal length lenses result in higher magnifications. Nevertheless, as long as high magnification of the field of view (test section) is achieved, even lenses with relatively short focal lengths can be used.

Spatial resolution of a BOS setup, which is a characterization of the blurred out-of-focus imaging, is influenced by two terms composed of several parameters, as given in equation (\ref{eq:delta_t}): it can be said that $\Delta_t$ and $f_{\#}$ are features of the camera and have a limited range of values, typically 5 -- 50 $\mu$m for $\Delta_t$, and 5.6 -- 32 for $f_{\#}$. For very low values of magnification, such as in the cases of BOS applied in outdoor measurements (natural backgrounds), the dominant term tends to be the second one. But, these cases also involve large distances between the background and the phase object, with $s_t$ usually of the order of a few meters. Hence, both terms would be accountable. For relatively large magnifications ($M > 0.1$), the second term can be neglected, by virtue of having a very low value for $\Delta_t$. Furthermore, when considering error propagation, even a small uncertainty in $s_t$, say 0.1\% (1 mm in a meter), would yield a larger uncertainty in $\delta_t$ than the value of $\Delta_t/M$ itself. Thus, a good approximation of the spatial resolution in terms of experimentally measured parameters can be expressed as
\begin{equation}\label{eq:simp-res}
\delta_t = \frac{s_i}{s_i+s_o}\,\frac{s_t}{f_{\#}}.
\end{equation}

This result is to be interpreted as follows: the angle of deflection $\varepsilon$ at the point $T$ is influenced by the variation of the refractive index $n(\mathbf{r})$ along the entire cone of light with the vertex at $B$ and base diameter $\delta_t$, i.e. $r = r_t \pm \delta_t$. Standard schlieren techniques use the phase object as the imaging object, for which $s_t = 0$. Hence, they have a superior spatial resolution as compared to BOS, because only the term $\Delta_t/M$ contributes to the uncertainty of measuring the angle of deflection.

\begin{figure}
\input{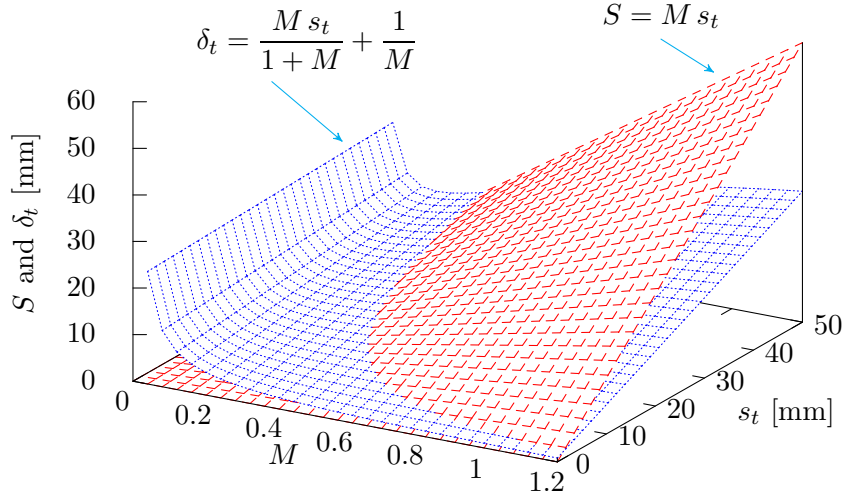}
\caption{BOS experiment design space.}
\label{fig:exp}
\end{figure}

Equations (\ref{eq:sens}) and (\ref{eq:delta_t}) are the main equations that characterize a setup for a BOS measurement, and their graphical presentation is given in figure \ref{fig:exp}. The surface plot for $\delta_t$ is shown for unit values of $\Delta_t$ and $f_{\#}$, and where $s_t<<s_o$. This figure presents the BOS experiment design space, meaning that once the acceptable range of values for resolution and sensitivity are decided, one can determine the distances between optical components when setting an experiment. 

This simple treatment of a BOS setup did not take into account several other factors that influence the measurement, such as lens aberrations, deviations from paraxial approximation, sagittal rays, and alike. But, some of these influences can be cancelled in the image evaluation stage, under the condition that the recording of the reference and measurement images is done with a fixed position of the camera and the test section relative to each other. Also, the irradiation model was very simple and did not consider the varying response of the image sensor. In addition, considering the small size of a pixel in image sensor, the diffraction effects may play a crucial role in imaging, especially if lasers are used for illumination.\footnote{In order to eliminate diffraction effects, equation (\ref{eq:delta_b}) with $k=0$ may serve as a threshold for the coherence length of the light source.} Nevertheless, the obtained results are quite general and can be applied to several variations of BOS setups. An important result is that neither $S$ or $\delta_t$ depend on the angle $\alpha$ (admittedly, as long as condition (\ref{eq:angles}) is satisfied).

\section{Conclusions}
\label{concl}
The present paper is a contribution in examining the performance of  background oriented schlieren (BOS), and it supplements our previous work in assessing BOS (Gojani and Obayashi 2012). Several guiding formulas for setting a BOS system are presented:
\begin{itemize}
\item detection limit is given by formula (\ref{eq:detection}),
\item condition for no-displacement of ray at the exit of test section is given by relation (\ref{eq:negl}),
\item dimensions of a field of view for a given spatial extent of a BOS setup is limited by relation (\ref{eq:field}),
\item system sensitivity can be estimated by equations (\ref{eq:sens}), 
\item the minimal diameter for a dot in the background is given by equation (\ref{eq:delta_b}), and
\item a fairly good approximation of spatial resolution from measurement setup is calculated by equation (\ref{eq:simp-res}).
\end{itemize}

These formulas should be used as a starting point in setting a BOS system and determining the needed instrumentation, but they are not conclusive for the latter, because in this analysis, the quantum efficiency of the image sensor and the temporal requirements of the experiment are not taken into account. The emphasis of this study, instead, was on the spatial setting of the instruments, so that desirable effects can be detected.

\begin{spacing}{1}

\end{spacing}

\end{document}

%% file: fig-bossetup.tex
\begin{tikzpicture}[gnuplot]
\gpsolidlines
\gpfill{color=gpbgfillcolor} (8.127,4.375)--(8.126,4.608)--(8.124,4.839)--(8.121,5.066)%
    --(8.116,5.289)--(8.109,5.504)--(8.102,5.710)--(8.093,5.907)--(8.083,6.092)%
    --(8.072,6.264)--(8.060,6.421)--(8.047,6.563)--(8.033,6.688)--(8.019,6.796)%
    --(8.004,6.885)--(7.988,6.955)--(7.972,7.005)--(7.956,7.036)--(7.940,7.046)%
    --(7.924,7.036)--(7.908,7.005)--(7.892,6.955)--(7.876,6.885)--(7.861,6.796)%
    --(7.847,6.688)--(7.833,6.563)--(7.820,6.421)--(7.808,6.264)--(7.797,6.092)%
    --(7.787,5.907)--(7.778,5.711)--(7.771,5.504)--(7.764,5.289)--(7.759,5.066)%
    --(7.756,4.839)--(7.754,4.608)--(7.753,4.375)--(7.754,4.142)--(7.756,3.911)%
    --(7.759,3.684)--(7.764,3.461)--(7.771,3.246)--(7.778,3.040)--(7.787,2.843)%
    --(7.797,2.658)--(7.808,2.486)--(7.820,2.329)--(7.833,2.187)--(7.847,2.062)%
    --(7.861,1.954)--(7.876,1.865)--(7.892,1.795)--(7.908,1.745)--(7.924,1.714)%
    --(7.940,1.704)--(7.956,1.714)--(7.972,1.745)--(7.988,1.795)--(8.004,1.865)%
    --(8.019,1.954)--(8.033,2.062)--(8.047,2.187)--(8.060,2.329)--(8.072,2.486)%
    --(8.083,2.658)--(8.093,2.843)--(8.102,3.039)--(8.109,3.246)--(8.116,3.461)%
    --(8.121,3.684)--(8.124,3.911)--(8.126,4.142)--cycle;
\gpcolor{gp lt color border}
\gpsetlinetype{gp lt plot 0}
\gpsetlinewidth{1.00}
\draw[gp path] (8.127,4.375)--(8.126,4.608)--(8.124,4.839)--(8.121,5.066)--(8.116,5.289)%
  --(8.109,5.504)--(8.102,5.710)--(8.093,5.907)--(8.083,6.092)--(8.072,6.264)--(8.060,6.421)%
  --(8.047,6.563)--(8.033,6.688)--(8.019,6.796)--(8.004,6.885)--(7.988,6.955)--(7.972,7.005)%
  --(7.956,7.036)--(7.940,7.046)--(7.924,7.036)--(7.908,7.005)--(7.892,6.955)--(7.876,6.885)%
  --(7.861,6.796)--(7.847,6.688)--(7.833,6.563)--(7.820,6.421)--(7.808,6.264)--(7.797,6.092)%
  --(7.787,5.907)--(7.778,5.711)--(7.771,5.504)--(7.764,5.289)--(7.759,5.066)--(7.756,4.839)%
  --(7.754,4.608)--(7.753,4.375)--(7.754,4.142)--(7.756,3.911)--(7.759,3.684)--(7.764,3.461)%
  --(7.771,3.246)--(7.778,3.040)--(7.787,2.843)--(7.797,2.658)--(7.808,2.486)--(7.820,2.329)%
  --(7.833,2.187)--(7.847,2.062)--(7.861,1.954)--(7.876,1.865)--(7.892,1.795)--(7.908,1.745)%
  --(7.924,1.714)--(7.940,1.704)--(7.956,1.714)--(7.972,1.745)--(7.988,1.795)--(8.004,1.865)%
  --(8.019,1.954)--(8.033,2.062)--(8.047,2.187)--(8.060,2.329)--(8.072,2.486)--(8.083,2.658)%
  --(8.093,2.843)--(8.102,3.039)--(8.109,3.246)--(8.116,3.461)--(8.121,3.684)--(8.124,3.911)%
  --(8.126,4.142)--cycle;
\gpfill{color=gpbgfillcolor} (3.701,1.704)--(4.699,1.704)--(4.699,7.046)--(3.701,7.046)--cycle;
\draw[gp path] (3.701,1.704)--(3.701,7.046)--(4.699,7.046)--(4.699,1.704)--cycle;
\gpfill{color=gpbgfillcolor} (1.769,2.238)--(1.769,2.243)--(1.768,2.248)--(1.767,2.254)%
    --(1.765,2.259)--(1.763,2.264)--(1.760,2.269)--(1.758,2.273)--(1.754,2.278)%
    --(1.751,2.282)--(1.747,2.285)--(1.742,2.289)--(1.738,2.291)--(1.733,2.294)%
    --(1.728,2.296)--(1.723,2.298)--(1.717,2.299)--(1.712,2.300)--(1.707,2.300)%
    --(1.701,2.300)--(1.696,2.299)--(1.690,2.298)--(1.685,2.296)--(1.680,2.294)%
    --(1.675,2.291)--(1.671,2.289)--(1.666,2.285)--(1.662,2.282)--(1.659,2.278)%
    --(1.655,2.273)--(1.653,2.269)--(1.650,2.264)--(1.648,2.259)--(1.646,2.254)%
    --(1.645,2.248)--(1.644,2.243)--(1.644,2.238)--(1.644,2.232)--(1.645,2.227)%
    --(1.646,2.221)--(1.648,2.216)--(1.650,2.211)--(1.653,2.206)--(1.655,2.202)%
    --(1.659,2.197)--(1.662,2.193)--(1.666,2.190)--(1.671,2.186)--(1.675,2.184)%
    --(1.680,2.181)--(1.685,2.179)--(1.690,2.177)--(1.696,2.176)--(1.701,2.175)%
    --(1.706,2.175)--(1.712,2.175)--(1.717,2.176)--(1.723,2.177)--(1.728,2.179)%
    --(1.733,2.181)--(1.738,2.184)--(1.742,2.186)--(1.747,2.190)--(1.751,2.193)%
    --(1.754,2.197)--(1.758,2.202)--(1.760,2.206)--(1.763,2.211)--(1.765,2.216)%
    --(1.767,2.221)--(1.768,2.227)--(1.769,2.232)--(1.769,2.237)--cycle;
\draw[gp path] (1.769,2.238)--(1.769,2.243)--(1.768,2.248)--(1.767,2.254)--(1.765,2.259)%
  --(1.763,2.264)--(1.760,2.269)--(1.758,2.273)--(1.754,2.278)--(1.751,2.282)--(1.747,2.285)%
  --(1.742,2.289)--(1.738,2.291)--(1.733,2.294)--(1.728,2.296)--(1.723,2.298)--(1.717,2.299)%
  --(1.712,2.300)--(1.707,2.300)--(1.701,2.300)--(1.696,2.299)--(1.690,2.298)--(1.685,2.296)%
  --(1.680,2.294)--(1.675,2.291)--(1.671,2.289)--(1.666,2.285)--(1.662,2.282)--(1.659,2.278)%
  --(1.655,2.273)--(1.653,2.269)--(1.650,2.264)--(1.648,2.259)--(1.646,2.254)--(1.645,2.248)%
  --(1.644,2.243)--(1.644,2.238)--(1.644,2.232)--(1.645,2.227)--(1.646,2.221)--(1.648,2.216)%
  --(1.650,2.211)--(1.653,2.206)--(1.655,2.202)--(1.659,2.197)--(1.662,2.193)--(1.666,2.190)%
  --(1.671,2.186)--(1.675,2.184)--(1.680,2.181)--(1.685,2.179)--(1.690,2.177)--(1.696,2.176)%
  --(1.701,2.175)--(1.706,2.175)--(1.712,2.175)--(1.717,2.176)--(1.723,2.177)--(1.728,2.179)%
  --(1.733,2.181)--(1.738,2.184)--(1.742,2.186)--(1.747,2.190)--(1.751,2.193)--(1.754,2.197)%
  --(1.758,2.202)--(1.760,2.206)--(1.763,2.211)--(1.765,2.216)--(1.767,2.221)--(1.768,2.227)%
  --(1.769,2.232)--(1.769,2.237);
\gpfill{color=gpbgfillcolor} (4.262,3.360)--(4.262,3.365)--(4.261,3.370)--(4.260,3.376)%
    --(4.258,3.381)--(4.256,3.386)--(4.253,3.391)--(4.251,3.395)--(4.247,3.400)%
    --(4.244,3.404)--(4.240,3.407)--(4.235,3.411)--(4.231,3.413)--(4.226,3.416)%
    --(4.221,3.418)--(4.216,3.420)--(4.210,3.421)--(4.205,3.422)--(4.200,3.422)%
    --(4.194,3.422)--(4.189,3.421)--(4.183,3.420)--(4.178,3.418)--(4.173,3.416)%
    --(4.168,3.413)--(4.164,3.411)--(4.159,3.407)--(4.155,3.404)--(4.152,3.400)%
    --(4.148,3.395)--(4.146,3.391)--(4.143,3.386)--(4.141,3.381)--(4.139,3.376)%
    --(4.138,3.370)--(4.137,3.365)--(4.137,3.360)--(4.137,3.354)--(4.138,3.349)%
    --(4.139,3.343)--(4.141,3.338)--(4.143,3.333)--(4.146,3.328)--(4.148,3.324)%
    --(4.152,3.319)--(4.155,3.315)--(4.159,3.312)--(4.164,3.308)--(4.168,3.306)%
    --(4.173,3.303)--(4.178,3.301)--(4.183,3.299)--(4.189,3.298)--(4.194,3.297)%
    --(4.199,3.297)--(4.205,3.297)--(4.210,3.298)--(4.216,3.299)--(4.221,3.301)%
    --(4.226,3.303)--(4.231,3.306)--(4.235,3.308)--(4.240,3.312)--(4.244,3.315)%
    --(4.247,3.319)--(4.251,3.324)--(4.253,3.328)--(4.256,3.333)--(4.258,3.338)%
    --(4.260,3.343)--(4.261,3.349)--(4.262,3.354)--(4.262,3.359)--cycle;
\draw[gp path] (4.262,3.360)--(4.262,3.365)--(4.261,3.370)--(4.260,3.376)--(4.258,3.381)%
  --(4.256,3.386)--(4.253,3.391)--(4.251,3.395)--(4.247,3.400)--(4.244,3.404)--(4.240,3.407)%
  --(4.235,3.411)--(4.231,3.413)--(4.226,3.416)--(4.221,3.418)--(4.216,3.420)--(4.210,3.421)%
  --(4.205,3.422)--(4.200,3.422)--(4.194,3.422)--(4.189,3.421)--(4.183,3.420)--(4.178,3.418)%
  --(4.173,3.416)--(4.168,3.413)--(4.164,3.411)--(4.159,3.407)--(4.155,3.404)--(4.152,3.400)%
  --(4.148,3.395)--(4.146,3.391)--(4.143,3.386)--(4.141,3.381)--(4.139,3.376)--(4.138,3.370)%
  --(4.137,3.365)--(4.137,3.360)--(4.137,3.354)--(4.138,3.349)--(4.139,3.343)--(4.141,3.338)%
  --(4.143,3.333)--(4.146,3.328)--(4.148,3.324)--(4.152,3.319)--(4.155,3.315)--(4.159,3.312)%
  --(4.164,3.308)--(4.168,3.306)--(4.173,3.303)--(4.178,3.301)--(4.183,3.299)--(4.189,3.298)%
  --(4.194,3.297)--(4.199,3.297)--(4.205,3.297)--(4.210,3.298)--(4.216,3.299)--(4.221,3.301)%
  --(4.226,3.303)--(4.231,3.306)--(4.235,3.308)--(4.240,3.312)--(4.244,3.315)--(4.247,3.319)%
  --(4.251,3.324)--(4.253,3.328)--(4.256,3.333)--(4.258,3.338)--(4.260,3.343)--(4.261,3.349)%
  --(4.262,3.354)--(4.262,3.359);
\gpfill{color=gpbgfillcolor} (9.872,4.375)--(9.872,4.380)--(9.871,4.385)--(9.870,4.391)%
    --(9.868,4.396)--(9.866,4.401)--(9.863,4.406)--(9.861,4.410)--(9.857,4.415)%
    --(9.854,4.419)--(9.850,4.422)--(9.845,4.426)--(9.841,4.428)--(9.836,4.431)%
    --(9.831,4.433)--(9.826,4.435)--(9.820,4.436)--(9.815,4.437)--(9.810,4.437)%
    --(9.804,4.437)--(9.799,4.436)--(9.793,4.435)--(9.788,4.433)--(9.783,4.431)%
    --(9.778,4.428)--(9.774,4.426)--(9.769,4.422)--(9.765,4.419)--(9.762,4.415)%
    --(9.758,4.410)--(9.756,4.406)--(9.753,4.401)--(9.751,4.396)--(9.749,4.391)%
    --(9.748,4.385)--(9.747,4.380)--(9.747,4.375)--(9.747,4.369)--(9.748,4.364)%
    --(9.749,4.358)--(9.751,4.353)--(9.753,4.348)--(9.756,4.343)--(9.758,4.339)%
    --(9.762,4.334)--(9.765,4.330)--(9.769,4.327)--(9.774,4.323)--(9.778,4.321)%
    --(9.783,4.318)--(9.788,4.316)--(9.793,4.314)--(9.799,4.313)--(9.804,4.312)%
    --(9.809,4.312)--(9.815,4.312)--(9.820,4.313)--(9.826,4.314)--(9.831,4.316)%
    --(9.836,4.318)--(9.841,4.321)--(9.845,4.323)--(9.850,4.327)--(9.854,4.330)%
    --(9.857,4.334)--(9.861,4.339)--(9.863,4.343)--(9.866,4.348)--(9.868,4.353)%
    --(9.870,4.358)--(9.871,4.364)--(9.872,4.369)--(9.872,4.374)--cycle;
\draw[gp path] (9.872,4.375)--(9.872,4.380)--(9.871,4.385)--(9.870,4.391)--(9.868,4.396)%
  --(9.866,4.401)--(9.863,4.406)--(9.861,4.410)--(9.857,4.415)--(9.854,4.419)--(9.850,4.422)%
  --(9.845,4.426)--(9.841,4.428)--(9.836,4.431)--(9.831,4.433)--(9.826,4.435)--(9.820,4.436)%
  --(9.815,4.437)--(9.810,4.437)--(9.804,4.437)--(9.799,4.436)--(9.793,4.435)--(9.788,4.433)%
  --(9.783,4.431)--(9.778,4.428)--(9.774,4.426)--(9.769,4.422)--(9.765,4.419)--(9.762,4.415)%
  --(9.758,4.410)--(9.756,4.406)--(9.753,4.401)--(9.751,4.396)--(9.749,4.391)--(9.748,4.385)%
  --(9.747,4.380)--(9.747,4.375)--(9.747,4.369)--(9.748,4.364)--(9.749,4.358)--(9.751,4.353)%
  --(9.753,4.348)--(9.756,4.343)--(9.758,4.339)--(9.762,4.334)--(9.765,4.330)--(9.769,4.327)%
  --(9.774,4.323)--(9.778,4.321)--(9.783,4.318)--(9.788,4.316)--(9.793,4.314)--(9.799,4.313)%
  --(9.804,4.312)--(9.809,4.312)--(9.815,4.312)--(9.820,4.313)--(9.826,4.314)--(9.831,4.316)%
  --(9.836,4.318)--(9.841,4.321)--(9.845,4.323)--(9.850,4.327)--(9.854,4.330)--(9.857,4.334)%
  --(9.861,4.339)--(9.863,4.343)--(9.866,4.348)--(9.868,4.353)--(9.870,4.358)--(9.871,4.364)%
  --(9.872,4.369)--(9.872,4.374);
\gpfill{color=gpbgfillcolor} (6.132,4.375)--(6.132,4.380)--(6.131,4.385)--(6.130,4.391)%
    --(6.128,4.396)--(6.126,4.401)--(6.123,4.406)--(6.121,4.410)--(6.117,4.415)%
    --(6.114,4.419)--(6.110,4.422)--(6.105,4.426)--(6.101,4.428)--(6.096,4.431)%
    --(6.091,4.433)--(6.086,4.435)--(6.080,4.436)--(6.075,4.437)--(6.070,4.437)%
    --(6.064,4.437)--(6.059,4.436)--(6.053,4.435)--(6.048,4.433)--(6.043,4.431)%
    --(6.038,4.428)--(6.034,4.426)--(6.029,4.422)--(6.025,4.419)--(6.022,4.415)%
    --(6.018,4.410)--(6.016,4.406)--(6.013,4.401)--(6.011,4.396)--(6.009,4.391)%
    --(6.008,4.385)--(6.007,4.380)--(6.007,4.375)--(6.007,4.369)--(6.008,4.364)%
    --(6.009,4.358)--(6.011,4.353)--(6.013,4.348)--(6.016,4.343)--(6.018,4.339)%
    --(6.022,4.334)--(6.025,4.330)--(6.029,4.327)--(6.034,4.323)--(6.038,4.321)%
    --(6.043,4.318)--(6.048,4.316)--(6.053,4.314)--(6.059,4.313)--(6.064,4.312)%
    --(6.069,4.312)--(6.075,4.312)--(6.080,4.313)--(6.086,4.314)--(6.091,4.316)%
    --(6.096,4.318)--(6.101,4.321)--(6.105,4.323)--(6.110,4.327)--(6.114,4.330)%
    --(6.117,4.334)--(6.121,4.339)--(6.123,4.343)--(6.126,4.348)--(6.128,4.353)%
    --(6.130,4.358)--(6.131,4.364)--(6.132,4.369)--(6.132,4.374)--cycle;
\draw[gp path] (6.132,4.375)--(6.132,4.380)--(6.131,4.385)--(6.130,4.391)--(6.128,4.396)%
  --(6.126,4.401)--(6.123,4.406)--(6.121,4.410)--(6.117,4.415)--(6.114,4.419)--(6.110,4.422)%
  --(6.105,4.426)--(6.101,4.428)--(6.096,4.431)--(6.091,4.433)--(6.086,4.435)--(6.080,4.436)%
  --(6.075,4.437)--(6.070,4.437)--(6.064,4.437)--(6.059,4.436)--(6.053,4.435)--(6.048,4.433)%
  --(6.043,4.431)--(6.038,4.428)--(6.034,4.426)--(6.029,4.422)--(6.025,4.419)--(6.022,4.415)%
  --(6.018,4.410)--(6.016,4.406)--(6.013,4.401)--(6.011,4.396)--(6.009,4.391)--(6.008,4.385)%
  --(6.007,4.380)--(6.007,4.375)--(6.007,4.369)--(6.008,4.364)--(6.009,4.358)--(6.011,4.353)%
  --(6.013,4.348)--(6.016,4.343)--(6.018,4.339)--(6.022,4.334)--(6.025,4.330)--(6.029,4.327)%
  --(6.034,4.323)--(6.038,4.321)--(6.043,4.318)--(6.048,4.316)--(6.053,4.314)--(6.059,4.313)%
  --(6.064,4.312)--(6.069,4.312)--(6.075,4.312)--(6.080,4.313)--(6.086,4.314)--(6.091,4.316)%
  --(6.096,4.318)--(6.101,4.321)--(6.105,4.323)--(6.110,4.327)--(6.114,4.330)--(6.117,4.334)%
  --(6.121,4.339)--(6.123,4.343)--(6.126,4.348)--(6.128,4.353)--(6.130,4.358)--(6.131,4.364)%
  --(6.132,4.369)--(6.132,4.374);
\gpfill{color=gpbgfillcolor} (10.673,5.290)--(10.673,5.295)--(10.672,5.300)--(10.671,5.306)%
    --(10.669,5.311)--(10.667,5.316)--(10.664,5.321)--(10.662,5.325)--(10.658,5.330)%
    --(10.655,5.334)--(10.651,5.337)--(10.646,5.341)--(10.642,5.343)--(10.637,5.346)%
    --(10.632,5.348)--(10.627,5.350)--(10.621,5.351)--(10.616,5.352)--(10.611,5.352)%
    --(10.605,5.352)--(10.600,5.351)--(10.594,5.350)--(10.589,5.348)--(10.584,5.346)%
    --(10.579,5.343)--(10.575,5.341)--(10.570,5.337)--(10.566,5.334)--(10.563,5.330)%
    --(10.559,5.325)--(10.557,5.321)--(10.554,5.316)--(10.552,5.311)--(10.550,5.306)%
    --(10.549,5.300)--(10.548,5.295)--(10.548,5.290)--(10.548,5.284)--(10.549,5.279)%
    --(10.550,5.273)--(10.552,5.268)--(10.554,5.263)--(10.557,5.258)--(10.559,5.254)%
    --(10.563,5.249)--(10.566,5.245)--(10.570,5.242)--(10.575,5.238)--(10.579,5.236)%
    --(10.584,5.233)--(10.589,5.231)--(10.594,5.229)--(10.600,5.228)--(10.605,5.227)%
    --(10.610,5.227)--(10.616,5.227)--(10.621,5.228)--(10.627,5.229)--(10.632,5.231)%
    --(10.637,5.233)--(10.642,5.236)--(10.646,5.238)--(10.651,5.242)--(10.655,5.245)%
    --(10.658,5.249)--(10.662,5.254)--(10.664,5.258)--(10.667,5.263)--(10.669,5.268)%
    --(10.671,5.273)--(10.672,5.279)--(10.673,5.284)--(10.673,5.289)--cycle;
\draw[gp path] (10.673,5.290)--(10.673,5.295)--(10.672,5.300)--(10.671,5.306)--(10.669,5.311)%
  --(10.667,5.316)--(10.664,5.321)--(10.662,5.325)--(10.658,5.330)--(10.655,5.334)--(10.651,5.337)%
  --(10.646,5.341)--(10.642,5.343)--(10.637,5.346)--(10.632,5.348)--(10.627,5.350)--(10.621,5.351)%
  --(10.616,5.352)--(10.611,5.352)--(10.605,5.352)--(10.600,5.351)--(10.594,5.350)--(10.589,5.348)%
  --(10.584,5.346)--(10.579,5.343)--(10.575,5.341)--(10.570,5.337)--(10.566,5.334)--(10.563,5.330)%
  --(10.559,5.325)--(10.557,5.321)--(10.554,5.316)--(10.552,5.311)--(10.550,5.306)--(10.549,5.300)%
  --(10.548,5.295)--(10.548,5.290)--(10.548,5.284)--(10.549,5.279)--(10.550,5.273)--(10.552,5.268)%
  --(10.554,5.263)--(10.557,5.258)--(10.559,5.254)--(10.563,5.249)--(10.566,5.245)--(10.570,5.242)%
  --(10.575,5.238)--(10.579,5.236)--(10.584,5.233)--(10.589,5.231)--(10.594,5.229)--(10.600,5.228)%
  --(10.605,5.227)--(10.610,5.227)--(10.616,5.227)--(10.621,5.228)--(10.627,5.229)--(10.632,5.231)%
  --(10.637,5.233)--(10.642,5.236)--(10.646,5.238)--(10.651,5.242)--(10.655,5.245)--(10.658,5.249)%
  --(10.662,5.254)--(10.664,5.258)--(10.667,5.263)--(10.669,5.268)--(10.671,5.273)--(10.672,5.279)%
  --(10.673,5.284)--(10.673,5.289);
\gpfill{color=gpbgfillcolor} (10.673,5.651)--(10.673,5.656)--(10.672,5.661)--(10.671,5.667)%
    --(10.669,5.672)--(10.667,5.677)--(10.664,5.682)--(10.662,5.686)--(10.658,5.691)%
    --(10.655,5.695)--(10.651,5.698)--(10.646,5.702)--(10.642,5.704)--(10.637,5.707)%
    --(10.632,5.709)--(10.627,5.711)--(10.621,5.712)--(10.616,5.713)--(10.611,5.713)%
    --(10.605,5.713)--(10.600,5.712)--(10.594,5.711)--(10.589,5.709)--(10.584,5.707)%
    --(10.579,5.704)--(10.575,5.702)--(10.570,5.698)--(10.566,5.695)--(10.563,5.691)%
    --(10.559,5.686)--(10.557,5.682)--(10.554,5.677)--(10.552,5.672)--(10.550,5.667)%
    --(10.549,5.661)--(10.548,5.656)--(10.548,5.651)--(10.548,5.645)--(10.549,5.640)%
    --(10.550,5.634)--(10.552,5.629)--(10.554,5.624)--(10.557,5.619)--(10.559,5.615)%
    --(10.563,5.610)--(10.566,5.606)--(10.570,5.603)--(10.575,5.599)--(10.579,5.597)%
    --(10.584,5.594)--(10.589,5.592)--(10.594,5.590)--(10.600,5.589)--(10.605,5.588)%
    --(10.610,5.588)--(10.616,5.588)--(10.621,5.589)--(10.627,5.590)--(10.632,5.592)%
    --(10.637,5.594)--(10.642,5.597)--(10.646,5.599)--(10.651,5.603)--(10.655,5.606)%
    --(10.658,5.610)--(10.662,5.615)--(10.664,5.619)--(10.667,5.624)--(10.669,5.629)%
    --(10.671,5.634)--(10.672,5.640)--(10.673,5.645)--(10.673,5.650)--cycle;
\draw[gp path] (10.673,5.651)--(10.673,5.656)--(10.672,5.661)--(10.671,5.667)--(10.669,5.672)%
  --(10.667,5.677)--(10.664,5.682)--(10.662,5.686)--(10.658,5.691)--(10.655,5.695)--(10.651,5.698)%
  --(10.646,5.702)--(10.642,5.704)--(10.637,5.707)--(10.632,5.709)--(10.627,5.711)--(10.621,5.712)%
  --(10.616,5.713)--(10.611,5.713)--(10.605,5.713)--(10.600,5.712)--(10.594,5.711)--(10.589,5.709)%
  --(10.584,5.707)--(10.579,5.704)--(10.575,5.702)--(10.570,5.698)--(10.566,5.695)--(10.563,5.691)%
  --(10.559,5.686)--(10.557,5.682)--(10.554,5.677)--(10.552,5.672)--(10.550,5.667)--(10.549,5.661)%
  --(10.548,5.656)--(10.548,5.651)--(10.548,5.645)--(10.549,5.640)--(10.550,5.634)--(10.552,5.629)%
  --(10.554,5.624)--(10.557,5.619)--(10.559,5.615)--(10.563,5.610)--(10.566,5.606)--(10.570,5.603)%
  --(10.575,5.599)--(10.579,5.597)--(10.584,5.594)--(10.589,5.592)--(10.594,5.590)--(10.600,5.589)%
  --(10.605,5.588)--(10.610,5.588)--(10.616,5.588)--(10.621,5.589)--(10.627,5.590)--(10.632,5.592)%
  --(10.637,5.594)--(10.642,5.597)--(10.646,5.599)--(10.651,5.603)--(10.655,5.606)--(10.658,5.610)%
  --(10.662,5.615)--(10.664,5.619)--(10.667,5.624)--(10.669,5.629)--(10.671,5.634)--(10.672,5.640)%
  --(10.673,5.645)--(10.673,5.650);
\node[gp node center] at (4.200,2.238) {$w$};
\node[gp node center] at (2.953,1.704) {$s_t$};
\node[gp node center] at (4.823,1.169) {$s_o$};
\node[gp node center] at (9.276,1.169) {$s_i$};
\node[gp node center] at (1.395,1.810) {$B$};
\node[gp node center] at (4.200,3.055) {$T$};
\node[gp node center] at (9.810,3.840) {$F$};
\node[gp node center] at (6.070,3.840) {$F$};
\node[gp node center,rotate=90] at (1.395,5.844) {Background};
\node[gp node center] at (4.200,7.313) {Test section};
\node[gp node center] at (4.200,6.778) {$n(\textbf{r})$};
\node[gp node center] at (7.940,7.313) {Lens};
\node[gp node center,rotate=-90] at (10.825,2.905) {Image sensor};
\node[gp node center] at (10.478,4.930) {$I$};
\node[gp node center] at (10.478,6.011) {$J$};
\node[gp node left] at (10.825,5.470) {$\Delta_{ij}$};
\node[gp node left] at (2.953,2.518) {$\alpha$};
\node[gp node center] at (7.1,5) {$\varepsilon$};
\gpsetlinetype{gp lt border}
\draw[gp path](1.707,1.704)--(1.707,7.046);
\draw[gp path](1.707,1.704)--(1.707,7.046);
\draw[gp path](10.611,1.704)--(10.611,7.046);
\gpcolor{gp lt color 4}
\gpsetlinetype{gp lt plot 0}
\draw[gp path,<->](4.699,1.971)--(3.701,1.971);
\draw[gp path,<->](4.200,1.436)--(1.707,1.436);
\draw[gp path,<->](7.940,0.902)--(1.707,0.902);
\draw[gp path,<->](10.611,0.902)--(7.940,0.902);
\draw[gp path,->](10.745,4.375)--(10.745,5.290);
\draw[gp path,->](10.745,6.109)--(10.745,5.651);
\draw[gp path,->](4.2,4.375)--(4.2,3.5);
\gpcolor{gp lt color border}
\gpsetlinetype{gp lt border}
\draw[gp path,->](2.953,2.238)--(2.829,2.743);
\draw[gp path](1.707,2.238)--(2.953,2.238);
\draw[gp path,->](7.005,4.622)--(6.693,5.323);
\gpcolor{gp lt color 1}
\gpsetlinetype{gp lt plot 1}
\draw[gp path, dotted] (0.460,4.375)--(0.576,4.375)--(0.692,4.375)--(0.808,4.375)--(0.924,4.375)%
  --(1.040,4.375)--(1.156,4.375)--(1.272,4.375)--(1.388,4.375)--(1.504,4.375)--(1.620,4.375)%
  --(1.736,4.375)--(1.852,4.375)--(1.968,4.375)--(2.084,4.375)--(2.200,4.375)--(2.316,4.375)%
  --(2.433,4.375)--(2.549,4.375)--(2.665,4.375)--(2.781,4.375)--(2.897,4.375)--(3.013,4.375)%
  --(3.129,4.375)--(3.245,4.375)--(3.361,4.375)--(3.477,4.375)--(3.593,4.375)--(3.709,4.375)%
  --(3.825,4.375)--(3.941,4.375)--(4.057,4.375)--(4.173,4.375)--(4.289,4.375)--(4.405,4.375)%
  --(4.521,4.375)--(4.637,4.375)--(4.753,4.375)--(4.869,4.375)--(4.985,4.375)--(5.101,4.375)%
  --(5.217,4.375)--(5.333,4.375)--(5.449,4.375)--(5.565,4.375)--(5.681,4.375)--(5.797,4.375)%
  --(5.913,4.375)--(6.029,4.375)--(6.145,4.375)--(6.262,4.375)--(6.378,4.375)--(6.494,4.375)%
  --(6.610,4.375)--(6.726,4.375)--(6.842,4.375)--(6.958,4.375)--(7.074,4.375)--(7.190,4.375)%
  --(7.306,4.375)--(7.422,4.375)--(7.538,4.375)--(7.654,4.375)--(7.770,4.375)--(7.886,4.375)%
  --(8.002,4.375)--(8.118,4.375)--(8.234,4.375)--(8.350,4.375)--(8.466,4.375)--(8.582,4.375)%
  --(8.698,4.375)--(8.814,4.375)--(8.930,4.375)--(9.046,4.375)--(9.162,4.375)--(9.278,4.375)%
  --(9.394,4.375)--(9.510,4.375)--(9.626,4.375)--(9.742,4.375)--(9.858,4.375)--(9.974,4.375)%
  --(10.091,4.375)--(10.207,4.375)--(10.323,4.375)--(10.439,4.375)--(10.555,4.375)--(10.671,4.375)%
  --(10.787,4.375)--(10.903,4.375)--(11.019,4.375)--(11.135,4.375)--(11.251,4.375)--(11.367,4.375)%
  --(11.483,4.375)--(11.599,4.375)--(11.715,4.375)--(11.831,4.375)--(11.947,4.375);
\gpcolor{gp lt color 0}
\gpsetlinetype{gp lt plot 0}
\gpsetlinewidth{2.00}
\draw[gp path] (4.289,3.430)--(4.405,3.521)--(4.521,3.612)--(4.637,3.704)--(4.753,3.795)%
  --(4.869,3.886)--(4.985,3.978)--(5.101,4.069)--(5.217,4.161)--(5.333,4.252)--(5.449,4.343)%
  --(5.565,4.435)--(5.681,4.526)--(5.797,4.617)--(5.913,4.709)--(6.029,4.800)--(6.145,4.892)%
  --(6.262,4.983)--(6.378,5.074)--(6.494,5.166)--(6.610,5.257)--(6.726,5.348)--(6.842,5.440)%
  --(6.958,5.531)--(7.074,5.622)--(7.190,5.714)--(7.306,5.805)--(7.422,5.897)--(7.538,5.988)%
  --(7.654,6.079)--(7.770,6.171)--(7.886,6.262)--(8.002,6.289)--(8.118,6.261)--(8.234,6.233)%
  --(8.350,6.204)--(8.466,6.176)--(8.582,6.147)--(8.698,6.119)--(8.814,6.091)--(8.930,6.062)%
  --(9.046,6.034)--(9.162,6.005)--(9.278,5.977)--(9.394,5.949)--(9.510,5.920)--(9.626,5.892)%
  --(9.742,5.863)--(9.858,5.835)--(9.974,5.807)--(10.091,5.778)--(10.207,5.750)--(10.323,5.721)%
  --(10.439,5.693)--(10.555,5.665);
\draw[gp path] (1.736,2.251)--(1.852,2.303)--(1.968,2.355)--(2.084,2.408)--(2.200,2.460)%
  --(2.316,2.512)--(2.433,2.564)--(2.549,2.616)--(2.665,2.669)--(2.781,2.721)--(2.897,2.773)%
  --(3.013,2.825)--(3.129,2.878)--(3.245,2.930)--(3.361,2.982)--(3.477,3.034)--(3.593,3.086)%
  --(3.709,3.139)--(3.825,3.191)--(3.941,3.243)--(4.057,3.295)--(4.173,3.347)--(4.289,3.400)%
  --(4.405,3.452)--(4.521,3.504)--(4.637,3.556)--(4.753,3.608)--(4.869,3.661)--(4.985,3.713)%
  --(5.101,3.765)--(5.217,3.817)--(5.333,3.869)--(5.449,3.922)--(5.565,3.974)--(5.681,4.026)%
  --(5.797,4.078)--(5.913,4.130)--(6.029,4.183)--(6.145,4.235)--(6.262,4.287)--(6.378,4.339)%
  --(6.494,4.391)--(6.610,4.444)--(6.726,4.496)--(6.842,4.548)--(6.958,4.600)--(7.074,4.653)%
  --(7.190,4.705)--(7.306,4.757)--(7.422,4.809)--(7.538,4.861)--(7.654,4.914)--(7.770,4.966)%
  --(7.886,5.018)--(8.002,5.048)--(8.118,5.059)--(8.234,5.070)--(8.350,5.080)--(8.466,5.091)%
  --(8.582,5.102)--(8.698,5.113)--(8.814,5.123)--(8.930,5.134)--(9.046,5.145)--(9.162,5.156)%
  --(9.278,5.166)--(9.394,5.177)--(9.510,5.188)--(9.626,5.199)--(9.742,5.210)--(9.858,5.220)%
  --(9.974,5.231)--(10.091,5.242)--(10.207,5.253)--(10.323,5.263)--(10.439,5.274)--(10.555,5.285);
\gpdefrectangularnode{gp plot 1}{\pgfpoint{0.460cm}{0.368cm}}{\pgfpoint{11.947cm}{8.381cm}}
\draw[thin,<-,black] (4.8,3.7) -- (5.2,4.0) to[out=33,in=90] (5.5,3.35);
\node[gp node center] at (5.5,3.055) {$\delta_e$};
\gpcolor{gp lt color 4}
\draw[thin, <->](7.94,5.1)--(7.94,6.2);
\node[gp node center] at (8.3,5.65) {$\delta_{\ell}$};
\draw[thin,->] (1.5,4.375) -- (1.5,2.238);
\node[gp node right] at (1.495,3.306) {$r_b$};
\node[gp node left] at (10.825,4.832) {$r_i$};
\node[gp node right] at (4.18,3.9) {$r_t$};
\end{tikzpicture}

%% file: fig-circleofconf2.tex
\begin{tikzpicture}[gnuplot]
\gpfill{color=gpbgfillcolor} (4.262,3.306)--(4.262,3.311)--(4.261,3.316)--(4.260,3.322)%
    --(4.258,3.327)--(4.256,3.332)--(4.253,3.337)--(4.251,3.341)--(4.247,3.346)%
    --(4.244,3.350)--(4.240,3.353)--(4.235,3.357)--(4.231,3.359)--(4.226,3.362)%
    --(4.221,3.364)--(4.216,3.366)--(4.210,3.367)--(4.205,3.368)--(4.200,3.368)%
    --(4.194,3.368)--(4.189,3.367)--(4.183,3.366)--(4.178,3.364)--(4.173,3.362)%
    --(4.168,3.359)--(4.164,3.357)--(4.159,3.353)--(4.155,3.350)--(4.152,3.346)%
    --(4.148,3.341)--(4.146,3.337)--(4.143,3.332)--(4.141,3.327)--(4.139,3.322)%
    --(4.138,3.316)--(4.137,3.311)--(4.137,3.306)--(4.137,3.300)--(4.138,3.295)%
    --(4.139,3.289)--(4.141,3.284)--(4.143,3.279)--(4.146,3.274)--(4.148,3.270)%
    --(4.152,3.265)--(4.155,3.261)--(4.159,3.258)--(4.164,3.254)--(4.168,3.252)%
    --(4.173,3.249)--(4.178,3.247)--(4.183,3.245)--(4.189,3.244)--(4.194,3.243)%
    --(4.199,3.243)--(4.205,3.243)--(4.210,3.244)--(4.216,3.245)--(4.221,3.247)%
    --(4.226,3.249)--(4.231,3.252)--(4.235,3.254)--(4.240,3.258)--(4.244,3.261)%
    --(4.247,3.265)--(4.251,3.270)--(4.253,3.274)--(4.256,3.279)--(4.258,3.284)%
    --(4.260,3.289)--(4.261,3.295)--(4.262,3.300)--(4.262,3.305)--cycle;
\gpcolor{gp lt color border}
\gpsetlinetype{gp lt plot 0}
\gpsetlinewidth{1.00}
\draw[gp path] (4.262,3.306)--(4.262,3.311)--(4.261,3.316)--(4.260,3.322)--(4.258,3.327)%
  --(4.256,3.332)--(4.253,3.337)--(4.251,3.341)--(4.247,3.346)--(4.244,3.350)--(4.240,3.353)%
  --(4.235,3.357)--(4.231,3.359)--(4.226,3.362)--(4.221,3.364)--(4.216,3.366)--(4.210,3.367)%
  --(4.205,3.368)--(4.200,3.368)--(4.194,3.368)--(4.189,3.367)--(4.183,3.366)--(4.178,3.364)%
  --(4.173,3.362)--(4.168,3.359)--(4.164,3.357)--(4.159,3.353)--(4.155,3.350)--(4.152,3.346)%
  --(4.148,3.341)--(4.146,3.337)--(4.143,3.332)--(4.141,3.327)--(4.139,3.322)--(4.138,3.316)%
  --(4.137,3.311)--(4.137,3.306)--(4.137,3.300)--(4.138,3.295)--(4.139,3.289)--(4.141,3.284)%
  --(4.143,3.279)--(4.146,3.274)--(4.148,3.270)--(4.152,3.265)--(4.155,3.261)--(4.159,3.258)%
  --(4.164,3.254)--(4.168,3.252)--(4.173,3.249)--(4.178,3.247)--(4.183,3.245)--(4.189,3.244)%
  --(4.194,3.243)--(4.199,3.243)--(4.205,3.243)--(4.210,3.244)--(4.216,3.245)--(4.221,3.247)%
  --(4.226,3.249)--(4.231,3.252)--(4.235,3.254)--(4.240,3.258)--(4.244,3.261)--(4.247,3.265)%
  --(4.251,3.270)--(4.253,3.274)--(4.256,3.279)--(4.258,3.284)--(4.260,3.289)--(4.261,3.295)%
  --(4.262,3.300)--(4.262,3.305);
\gpfill{color=gpbgfillcolor} (9.872,4.375)--(9.872,4.380)--(9.871,4.385)--(9.870,4.391)%
    --(9.868,4.396)--(9.866,4.401)--(9.863,4.406)--(9.861,4.410)--(9.857,4.415)%
    --(9.854,4.419)--(9.850,4.422)--(9.845,4.426)--(9.841,4.428)--(9.836,4.431)%
    --(9.831,4.433)--(9.826,4.435)--(9.820,4.436)--(9.815,4.437)--(9.810,4.437)%
    --(9.804,4.437)--(9.799,4.436)--(9.793,4.435)--(9.788,4.433)--(9.783,4.431)%
    --(9.778,4.428)--(9.774,4.426)--(9.769,4.422)--(9.765,4.419)--(9.762,4.415)%
    --(9.758,4.410)--(9.756,4.406)--(9.753,4.401)--(9.751,4.396)--(9.749,4.391)%
    --(9.748,4.385)--(9.747,4.380)--(9.747,4.375)--(9.747,4.369)--(9.748,4.364)%
    --(9.749,4.358)--(9.751,4.353)--(9.753,4.348)--(9.756,4.343)--(9.758,4.339)%
    --(9.762,4.334)--(9.765,4.330)--(9.769,4.327)--(9.774,4.323)--(9.778,4.321)%
    --(9.783,4.318)--(9.788,4.316)--(9.793,4.314)--(9.799,4.313)--(9.804,4.312)%
    --(9.809,4.312)--(9.815,4.312)--(9.820,4.313)--(9.826,4.314)--(9.831,4.316)%
    --(9.836,4.318)--(9.841,4.321)--(9.845,4.323)--(9.850,4.327)--(9.854,4.330)%
    --(9.857,4.334)--(9.861,4.339)--(9.863,4.343)--(9.866,4.348)--(9.868,4.353)%
    --(9.870,4.358)--(9.871,4.364)--(9.872,4.369)--(9.872,4.374)--cycle;
\draw[gp path] (9.872,4.375)--(9.872,4.380)--(9.871,4.385)--(9.870,4.391)--(9.868,4.396)%
  --(9.866,4.401)--(9.863,4.406)--(9.861,4.410)--(9.857,4.415)--(9.854,4.419)--(9.850,4.422)%
  --(9.845,4.426)--(9.841,4.428)--(9.836,4.431)--(9.831,4.433)--(9.826,4.435)--(9.820,4.436)%
  --(9.815,4.437)--(9.810,4.437)--(9.804,4.437)--(9.799,4.436)--(9.793,4.435)--(9.788,4.433)%
  --(9.783,4.431)--(9.778,4.428)--(9.774,4.426)--(9.769,4.422)--(9.765,4.419)--(9.762,4.415)%
  --(9.758,4.410)--(9.756,4.406)--(9.753,4.401)--(9.751,4.396)--(9.749,4.391)--(9.748,4.385)%
  --(9.747,4.380)--(9.747,4.375)--(9.747,4.369)--(9.748,4.364)--(9.749,4.358)--(9.751,4.353)%
  --(9.753,4.348)--(9.756,4.343)--(9.758,4.339)--(9.762,4.334)--(9.765,4.330)--(9.769,4.327)%
  --(9.774,4.323)--(9.778,4.321)--(9.783,4.318)--(9.788,4.316)--(9.793,4.314)--(9.799,4.313)%
  --(9.804,4.312)--(9.809,4.312)--(9.815,4.312)--(9.820,4.313)--(9.826,4.314)--(9.831,4.316)%
  --(9.836,4.318)--(9.841,4.321)--(9.845,4.323)--(9.850,4.327)--(9.854,4.330)--(9.857,4.334)%
  --(9.861,4.339)--(9.863,4.343)--(9.866,4.348)--(9.868,4.353)--(9.870,4.358)--(9.871,4.364)%
  --(9.872,4.369)--(9.872,4.374);
\gpfill{color=gpbgfillcolor} (6.132,4.375)--(6.132,4.380)--(6.131,4.385)--(6.130,4.391)%
    --(6.128,4.396)--(6.126,4.401)--(6.123,4.406)--(6.121,4.410)--(6.117,4.415)%
    --(6.114,4.419)--(6.110,4.422)--(6.105,4.426)--(6.101,4.428)--(6.096,4.431)%
    --(6.091,4.433)--(6.086,4.435)--(6.080,4.436)--(6.075,4.437)--(6.070,4.437)%
    --(6.064,4.437)--(6.059,4.436)--(6.053,4.435)--(6.048,4.433)--(6.043,4.431)%
    --(6.038,4.428)--(6.034,4.426)--(6.029,4.422)--(6.025,4.419)--(6.022,4.415)%
    --(6.018,4.410)--(6.016,4.406)--(6.013,4.401)--(6.011,4.396)--(6.009,4.391)%
    --(6.008,4.385)--(6.007,4.380)--(6.007,4.375)--(6.007,4.369)--(6.008,4.364)%
    --(6.009,4.358)--(6.011,4.353)--(6.013,4.348)--(6.016,4.343)--(6.018,4.339)%
    --(6.022,4.334)--(6.025,4.330)--(6.029,4.327)--(6.034,4.323)--(6.038,4.321)%
    --(6.043,4.318)--(6.048,4.316)--(6.053,4.314)--(6.059,4.313)--(6.064,4.312)%
    --(6.069,4.312)--(6.075,4.312)--(6.080,4.313)--(6.086,4.314)--(6.091,4.316)%
    --(6.096,4.318)--(6.101,4.321)--(6.105,4.323)--(6.110,4.327)--(6.114,4.330)%
    --(6.117,4.334)--(6.121,4.339)--(6.123,4.343)--(6.126,4.348)--(6.128,4.353)%
    --(6.130,4.358)--(6.131,4.364)--(6.132,4.369)--(6.132,4.374)--cycle;
\draw[gp path] (6.132,4.375)--(6.132,4.380)--(6.131,4.385)--(6.130,4.391)--(6.128,4.396)%
  --(6.126,4.401)--(6.123,4.406)--(6.121,4.410)--(6.117,4.415)--(6.114,4.419)--(6.110,4.422)%
  --(6.105,4.426)--(6.101,4.428)--(6.096,4.431)--(6.091,4.433)--(6.086,4.435)--(6.080,4.436)%
  --(6.075,4.437)--(6.070,4.437)--(6.064,4.437)--(6.059,4.436)--(6.053,4.435)--(6.048,4.433)%
  --(6.043,4.431)--(6.038,4.428)--(6.034,4.426)--(6.029,4.422)--(6.025,4.419)--(6.022,4.415)%
  --(6.018,4.410)--(6.016,4.406)--(6.013,4.401)--(6.011,4.396)--(6.009,4.391)--(6.008,4.385)%
  --(6.007,4.380)--(6.007,4.375)--(6.007,4.369)--(6.008,4.364)--(6.009,4.358)--(6.011,4.353)%
  --(6.013,4.348)--(6.016,4.343)--(6.018,4.339)--(6.022,4.334)--(6.025,4.330)--(6.029,4.327)%
  --(6.034,4.323)--(6.038,4.321)--(6.043,4.318)--(6.048,4.316)--(6.053,4.314)--(6.059,4.313)%
  --(6.064,4.312)--(6.069,4.312)--(6.075,4.312)--(6.080,4.313)--(6.086,4.314)--(6.091,4.316)%
  --(6.096,4.318)--(6.101,4.321)--(6.105,4.323)--(6.110,4.327)--(6.114,4.330)--(6.117,4.334)%
  --(6.121,4.339)--(6.123,4.343)--(6.126,4.348)--(6.128,4.353)--(6.130,4.358)--(6.131,4.364)%
  --(6.132,4.369)--(6.132,4.374);
\gpfill{color=gpbgfillcolor} (10.673,5.519)--(10.673,5.524)--(10.672,5.529)--(10.671,5.535)%
    --(10.669,5.540)--(10.667,5.545)--(10.664,5.550)--(10.662,5.554)--(10.658,5.559)%
    --(10.655,5.563)--(10.651,5.566)--(10.646,5.570)--(10.642,5.572)--(10.637,5.575)%
    --(10.632,5.577)--(10.627,5.579)--(10.621,5.580)--(10.616,5.581)--(10.611,5.581)%
    --(10.605,5.581)--(10.600,5.580)--(10.594,5.579)--(10.589,5.577)--(10.584,5.575)%
    --(10.579,5.572)--(10.575,5.570)--(10.570,5.566)--(10.566,5.563)--(10.563,5.559)%
    --(10.559,5.554)--(10.557,5.550)--(10.554,5.545)--(10.552,5.540)--(10.550,5.535)%
    --(10.549,5.529)--(10.548,5.524)--(10.548,5.519)--(10.548,5.513)--(10.549,5.508)%
    --(10.550,5.502)--(10.552,5.497)--(10.554,5.492)--(10.557,5.487)--(10.559,5.483)%
    --(10.563,5.478)--(10.566,5.474)--(10.570,5.471)--(10.575,5.467)--(10.579,5.465)%
    --(10.584,5.462)--(10.589,5.460)--(10.594,5.458)--(10.600,5.457)--(10.605,5.456)%
    --(10.610,5.456)--(10.616,5.456)--(10.621,5.457)--(10.627,5.458)--(10.632,5.460)%
    --(10.637,5.462)--(10.642,5.465)--(10.646,5.467)--(10.651,5.471)--(10.655,5.474)%
    --(10.658,5.478)--(10.662,5.483)--(10.664,5.487)--(10.667,5.492)--(10.669,5.497)%
    --(10.671,5.502)--(10.672,5.508)--(10.673,5.513)--(10.673,5.518)--cycle;
\draw[gp path] (10.673,5.519)--(10.673,5.524)--(10.672,5.529)--(10.671,5.535)--(10.669,5.540)%
  --(10.667,5.545)--(10.664,5.550)--(10.662,5.554)--(10.658,5.559)--(10.655,5.563)--(10.651,5.566)%
  --(10.646,5.570)--(10.642,5.572)--(10.637,5.575)--(10.632,5.577)--(10.627,5.579)--(10.621,5.580)%
  --(10.616,5.581)--(10.611,5.581)--(10.605,5.581)--(10.600,5.580)--(10.594,5.579)--(10.589,5.577)%
  --(10.584,5.575)--(10.579,5.572)--(10.575,5.570)--(10.570,5.566)--(10.566,5.563)--(10.563,5.559)%
  --(10.559,5.554)--(10.557,5.550)--(10.554,5.545)--(10.552,5.540)--(10.550,5.535)--(10.549,5.529)%
  --(10.548,5.524)--(10.548,5.519)--(10.548,5.513)--(10.549,5.508)--(10.550,5.502)--(10.552,5.497)%
  --(10.554,5.492)--(10.557,5.487)--(10.559,5.483)--(10.563,5.478)--(10.566,5.474)--(10.570,5.471)%
  --(10.575,5.467)--(10.579,5.465)--(10.584,5.462)--(10.589,5.460)--(10.594,5.458)--(10.600,5.457)%
  --(10.605,5.456)--(10.610,5.456)--(10.616,5.456)--(10.621,5.457)--(10.627,5.458)--(10.632,5.460)%
  --(10.637,5.462)--(10.642,5.465)--(10.646,5.467)--(10.651,5.471)--(10.655,5.474)--(10.658,5.478)%
  --(10.662,5.483)--(10.664,5.487)--(10.667,5.492)--(10.669,5.497)--(10.671,5.502)--(10.672,5.508)%
  --(10.673,5.513)--(10.673,5.518);
\gpfill{color=gpbgfillcolor} (10.673,4.756)--(10.673,4.761)--(10.672,4.766)--(10.671,4.772)%
    --(10.669,4.777)--(10.667,4.782)--(10.664,4.787)--(10.662,4.791)--(10.658,4.796)%
    --(10.655,4.800)--(10.651,4.803)--(10.646,4.807)--(10.642,4.809)--(10.637,4.812)%
    --(10.632,4.814)--(10.627,4.816)--(10.621,4.817)--(10.616,4.818)--(10.611,4.818)%
    --(10.605,4.818)--(10.600,4.817)--(10.594,4.816)--(10.589,4.814)--(10.584,4.812)%
    --(10.579,4.809)--(10.575,4.807)--(10.570,4.803)--(10.566,4.800)--(10.563,4.796)%
    --(10.559,4.791)--(10.557,4.787)--(10.554,4.782)--(10.552,4.777)--(10.550,4.772)%
    --(10.549,4.766)--(10.548,4.761)--(10.548,4.756)--(10.548,4.750)--(10.549,4.745)%
    --(10.550,4.739)--(10.552,4.734)--(10.554,4.729)--(10.557,4.724)--(10.559,4.720)%
    --(10.563,4.715)--(10.566,4.711)--(10.570,4.708)--(10.575,4.704)--(10.579,4.702)%
    --(10.584,4.699)--(10.589,4.697)--(10.594,4.695)--(10.600,4.694)--(10.605,4.693)%
    --(10.610,4.693)--(10.616,4.693)--(10.621,4.694)--(10.627,4.695)--(10.632,4.697)%
    --(10.637,4.699)--(10.642,4.702)--(10.646,4.704)--(10.651,4.708)--(10.655,4.711)%
    --(10.658,4.715)--(10.662,4.720)--(10.664,4.724)--(10.667,4.729)--(10.669,4.734)%
    --(10.671,4.739)--(10.672,4.745)--(10.673,4.750)--(10.673,4.755)--cycle;
\draw[gp path] (10.673,4.756)--(10.673,4.761)--(10.672,4.766)--(10.671,4.772)--(10.669,4.777)%
  --(10.667,4.782)--(10.664,4.787)--(10.662,4.791)--(10.658,4.796)--(10.655,4.800)--(10.651,4.803)%
  --(10.646,4.807)--(10.642,4.809)--(10.637,4.812)--(10.632,4.814)--(10.627,4.816)--(10.621,4.817)%
  --(10.616,4.818)--(10.611,4.818)--(10.605,4.818)--(10.600,4.817)--(10.594,4.816)--(10.589,4.814)%
  --(10.584,4.812)--(10.579,4.809)--(10.575,4.807)--(10.570,4.803)--(10.566,4.800)--(10.563,4.796)%
  --(10.559,4.791)--(10.557,4.787)--(10.554,4.782)--(10.552,4.777)--(10.550,4.772)--(10.549,4.766)%
  --(10.548,4.761)--(10.548,4.756)--(10.548,4.750)--(10.549,4.745)--(10.550,4.739)--(10.552,4.734)%
  --(10.554,4.729)--(10.557,4.724)--(10.559,4.720)--(10.563,4.715)--(10.566,4.711)--(10.570,4.708)%
  --(10.575,4.704)--(10.579,4.702)--(10.584,4.699)--(10.589,4.697)--(10.594,4.695)--(10.600,4.694)%
  --(10.605,4.693)--(10.610,4.693)--(10.616,4.693)--(10.621,4.694)--(10.627,4.695)--(10.632,4.697)%
  --(10.637,4.699)--(10.642,4.702)--(10.646,4.704)--(10.651,4.708)--(10.655,4.711)--(10.658,4.715)%
  --(10.662,4.720)--(10.664,4.724)--(10.667,4.729)--(10.669,4.734)--(10.671,4.739)--(10.672,4.745)%
  --(10.673,4.750)--(10.673,4.755);
\gpfill{color=gpbgfillcolor} (1.769,1.704)--(1.769,1.709)--(1.768,1.714)--(1.767,1.720)%
    --(1.765,1.725)--(1.763,1.730)--(1.760,1.735)--(1.758,1.739)--(1.754,1.744)%
    --(1.751,1.748)--(1.747,1.751)--(1.742,1.755)--(1.738,1.757)--(1.733,1.760)%
    --(1.728,1.762)--(1.723,1.764)--(1.717,1.765)--(1.712,1.766)--(1.707,1.766)%
    --(1.701,1.766)--(1.696,1.765)--(1.690,1.764)--(1.685,1.762)--(1.680,1.760)%
    --(1.675,1.757)--(1.671,1.755)--(1.666,1.751)--(1.662,1.748)--(1.659,1.744)%
    --(1.655,1.739)--(1.653,1.735)--(1.650,1.730)--(1.648,1.725)--(1.646,1.720)%
    --(1.645,1.714)--(1.644,1.709)--(1.644,1.704)--(1.644,1.698)--(1.645,1.693)%
    --(1.646,1.687)--(1.648,1.682)--(1.650,1.677)--(1.653,1.672)--(1.655,1.668)%
    --(1.659,1.663)--(1.662,1.659)--(1.666,1.656)--(1.671,1.652)--(1.675,1.650)%
    --(1.680,1.647)--(1.685,1.645)--(1.690,1.643)--(1.696,1.642)--(1.701,1.641)%
    --(1.706,1.641)--(1.712,1.641)--(1.717,1.642)--(1.723,1.643)--(1.728,1.645)%
    --(1.733,1.647)--(1.738,1.650)--(1.742,1.652)--(1.747,1.656)--(1.751,1.659)%
    --(1.754,1.663)--(1.758,1.668)--(1.760,1.672)--(1.763,1.677)--(1.765,1.682)%
    --(1.767,1.687)--(1.768,1.693)--(1.769,1.698)--(1.769,1.703)--cycle;
\draw[gp path] (1.769,1.704)--(1.769,1.709)--(1.768,1.714)--(1.767,1.720)--(1.765,1.725)%
  --(1.763,1.730)--(1.760,1.735)--(1.758,1.739)--(1.754,1.744)--(1.751,1.748)--(1.747,1.751)%
  --(1.742,1.755)--(1.738,1.757)--(1.733,1.760)--(1.728,1.762)--(1.723,1.764)--(1.717,1.765)%
  --(1.712,1.766)--(1.707,1.766)--(1.701,1.766)--(1.696,1.765)--(1.690,1.764)--(1.685,1.762)%
  --(1.680,1.760)--(1.675,1.757)--(1.671,1.755)--(1.666,1.751)--(1.662,1.748)--(1.659,1.744)%
  --(1.655,1.739)--(1.653,1.735)--(1.650,1.730)--(1.648,1.725)--(1.646,1.720)--(1.645,1.714)%
  --(1.644,1.709)--(1.644,1.704)--(1.644,1.698)--(1.645,1.693)--(1.646,1.687)--(1.648,1.682)%
  --(1.650,1.677)--(1.653,1.672)--(1.655,1.668)--(1.659,1.663)--(1.662,1.659)--(1.666,1.656)%
  --(1.671,1.652)--(1.675,1.650)--(1.680,1.647)--(1.685,1.645)--(1.690,1.643)--(1.696,1.642)%
  --(1.701,1.641)--(1.706,1.641)--(1.712,1.641)--(1.717,1.642)--(1.723,1.643)--(1.728,1.645)%
  --(1.733,1.647)--(1.738,1.650)--(1.742,1.652)--(1.747,1.656)--(1.751,1.659)--(1.754,1.663)%
  --(1.758,1.668)--(1.760,1.672)--(1.763,1.677)--(1.765,1.682)--(1.767,1.687)--(1.768,1.693)%
  --(1.769,1.698)--(1.769,1.703);
\gpfill{color=gpbgfillcolor} (1.769,3.484)--(1.769,3.489)--(1.768,3.494)--(1.767,3.500)%
    --(1.765,3.505)--(1.763,3.510)--(1.760,3.515)--(1.758,3.519)--(1.754,3.524)%
    --(1.751,3.528)--(1.747,3.531)--(1.742,3.535)--(1.738,3.537)--(1.733,3.540)%
    --(1.728,3.542)--(1.723,3.544)--(1.717,3.545)--(1.712,3.546)--(1.707,3.546)%
    --(1.701,3.546)--(1.696,3.545)--(1.690,3.544)--(1.685,3.542)--(1.680,3.540)%
    --(1.675,3.537)--(1.671,3.535)--(1.666,3.531)--(1.662,3.528)--(1.659,3.524)%
    --(1.655,3.519)--(1.653,3.515)--(1.650,3.510)--(1.648,3.505)--(1.646,3.500)%
    --(1.645,3.494)--(1.644,3.489)--(1.644,3.484)--(1.644,3.478)--(1.645,3.473)%
    --(1.646,3.467)--(1.648,3.462)--(1.650,3.457)--(1.653,3.452)--(1.655,3.448)%
    --(1.659,3.443)--(1.662,3.439)--(1.666,3.436)--(1.671,3.432)--(1.675,3.430)%
    --(1.680,3.427)--(1.685,3.425)--(1.690,3.423)--(1.696,3.422)--(1.701,3.421)%
    --(1.706,3.421)--(1.712,3.421)--(1.717,3.422)--(1.723,3.423)--(1.728,3.425)%
    --(1.733,3.427)--(1.738,3.430)--(1.742,3.432)--(1.747,3.436)--(1.751,3.439)%
    --(1.754,3.443)--(1.758,3.448)--(1.760,3.452)--(1.763,3.457)--(1.765,3.462)%
    --(1.767,3.467)--(1.768,3.473)--(1.769,3.478)--(1.769,3.483)--cycle;
\draw[gp path] (1.769,3.484)--(1.769,3.489)--(1.768,3.494)--(1.767,3.500)--(1.765,3.505)%
  --(1.763,3.510)--(1.760,3.515)--(1.758,3.519)--(1.754,3.524)--(1.751,3.528)--(1.747,3.531)%
  --(1.742,3.535)--(1.738,3.537)--(1.733,3.540)--(1.728,3.542)--(1.723,3.544)--(1.717,3.545)%
  --(1.712,3.546)--(1.707,3.546)--(1.701,3.546)--(1.696,3.545)--(1.690,3.544)--(1.685,3.542)%
  --(1.680,3.540)--(1.675,3.537)--(1.671,3.535)--(1.666,3.531)--(1.662,3.528)--(1.659,3.524)%
  --(1.655,3.519)--(1.653,3.515)--(1.650,3.510)--(1.648,3.505)--(1.646,3.500)--(1.645,3.494)%
  --(1.644,3.489)--(1.644,3.484)--(1.644,3.478)--(1.645,3.473)--(1.646,3.467)--(1.648,3.462)%
  --(1.650,3.457)--(1.653,3.452)--(1.655,3.448)--(1.659,3.443)--(1.662,3.439)--(1.666,3.436)%
  --(1.671,3.432)--(1.675,3.430)--(1.680,3.427)--(1.685,3.425)--(1.690,3.423)--(1.696,3.422)%
  --(1.701,3.421)--(1.706,3.421)--(1.712,3.421)--(1.717,3.422)--(1.723,3.423)--(1.728,3.425)%
  --(1.733,3.427)--(1.738,3.430)--(1.742,3.432)--(1.747,3.436)--(1.751,3.439)--(1.754,3.443)%
  --(1.758,3.448)--(1.760,3.452)--(1.763,3.457)--(1.765,3.462)--(1.767,3.467)--(1.768,3.473)%
  --(1.769,3.478)--(1.769,3.483);
\gpfill{color=gpbgfillcolor} (1.769,2.594)--(1.769,2.599)--(1.768,2.604)--(1.767,2.610)%
    --(1.765,2.615)--(1.763,2.620)--(1.760,2.625)--(1.758,2.629)--(1.754,2.634)%
    --(1.751,2.638)--(1.747,2.641)--(1.742,2.645)--(1.738,2.647)--(1.733,2.650)%
    --(1.728,2.652)--(1.723,2.654)--(1.717,2.655)--(1.712,2.656)--(1.707,2.656)%
    --(1.701,2.656)--(1.696,2.655)--(1.690,2.654)--(1.685,2.652)--(1.680,2.650)%
    --(1.675,2.647)--(1.671,2.645)--(1.666,2.641)--(1.662,2.638)--(1.659,2.634)%
    --(1.655,2.629)--(1.653,2.625)--(1.650,2.620)--(1.648,2.615)--(1.646,2.610)%
    --(1.645,2.604)--(1.644,2.599)--(1.644,2.594)--(1.644,2.588)--(1.645,2.583)%
    --(1.646,2.577)--(1.648,2.572)--(1.650,2.567)--(1.653,2.562)--(1.655,2.558)%
    --(1.659,2.553)--(1.662,2.549)--(1.666,2.546)--(1.671,2.542)--(1.675,2.540)%
    --(1.680,2.537)--(1.685,2.535)--(1.690,2.533)--(1.696,2.532)--(1.701,2.531)%
    --(1.706,2.531)--(1.712,2.531)--(1.717,2.532)--(1.723,2.533)--(1.728,2.535)%
    --(1.733,2.537)--(1.738,2.540)--(1.742,2.542)--(1.747,2.546)--(1.751,2.549)%
    --(1.754,2.553)--(1.758,2.558)--(1.760,2.562)--(1.763,2.567)--(1.765,2.572)%
    --(1.767,2.577)--(1.768,2.583)--(1.769,2.588)--(1.769,2.593)--cycle;
\draw[gp path] (1.769,2.594)--(1.769,2.599)--(1.768,2.604)--(1.767,2.610)--(1.765,2.615)%
  --(1.763,2.620)--(1.760,2.625)--(1.758,2.629)--(1.754,2.634)--(1.751,2.638)--(1.747,2.641)%
  --(1.742,2.645)--(1.738,2.647)--(1.733,2.650)--(1.728,2.652)--(1.723,2.654)--(1.717,2.655)%
  --(1.712,2.656)--(1.707,2.656)--(1.701,2.656)--(1.696,2.655)--(1.690,2.654)--(1.685,2.652)%
  --(1.680,2.650)--(1.675,2.647)--(1.671,2.645)--(1.666,2.641)--(1.662,2.638)--(1.659,2.634)%
  --(1.655,2.629)--(1.653,2.625)--(1.650,2.620)--(1.648,2.615)--(1.646,2.610)--(1.645,2.604)%
  --(1.644,2.599)--(1.644,2.594)--(1.644,2.588)--(1.645,2.583)--(1.646,2.577)--(1.648,2.572)%
  --(1.650,2.567)--(1.653,2.562)--(1.655,2.558)--(1.659,2.553)--(1.662,2.549)--(1.666,2.546)%
  --(1.671,2.542)--(1.675,2.540)--(1.680,2.537)--(1.685,2.535)--(1.690,2.533)--(1.696,2.532)%
  --(1.701,2.531)--(1.706,2.531)--(1.712,2.531)--(1.717,2.532)--(1.723,2.533)--(1.728,2.535)%
  --(1.733,2.537)--(1.738,2.540)--(1.742,2.542)--(1.747,2.546)--(1.751,2.549)--(1.754,2.553)%
  --(1.758,2.558)--(1.760,2.562)--(1.763,2.567)--(1.765,2.572)--(1.767,2.577)--(1.768,2.583)%
  --(1.769,2.588)--(1.769,2.593);
\gpfill{color=gpbgfillcolor} (10.638,4.428)--(10.691,4.428)--(10.691,4.855)--(10.638,4.855)--cycle;
\draw[gp path] (10.638,4.428)--(10.638,4.855)--(10.691,4.855)--(10.691,4.428)--cycle;
\gpfill{color=gpbgfillcolor} (10.638,4.962)--(10.691,4.962)--(10.691,5.389)--(10.638,5.389)--cycle;
\draw[gp path] (10.638,4.962)--(10.638,5.389)--(10.691,5.389)--(10.691,4.962)--cycle;
\gpfill{color=gpbgfillcolor} (10.638,5.496)--(10.691,5.496)--(10.691,5.924)--(10.638,5.924)--cycle;
\draw[gp path] (10.638,5.496)--(10.638,5.924)--(10.691,5.924)--(10.691,5.496)--cycle;
\gpfill{color=gpbgfillcolor} (10.638,6.031)--(10.691,6.031)--(10.691,6.458)--(10.638,6.458)--cycle;
\draw[gp path] (10.638,6.031)--(10.638,6.458)--(10.691,6.458)--(10.691,6.031)--cycle;
\gpfill{color=gpbgfillcolor} (10.638,6.565)--(10.691,6.565)--(10.691,6.992)--(10.638,6.992)--cycle;
\draw[gp path] (10.638,6.565)--(10.638,6.992)--(10.691,6.992)--(10.691,6.565)--cycle;
\gpfill{color=gpbgfillcolor} (10.638,3.894)--(10.691,3.894)--(10.691,4.321)--(10.638,4.321)--cycle;
\draw[gp path] (10.638,3.894)--(10.638,4.321)--(10.691,4.321)--(10.691,3.894)--cycle;
\gpfill{color=gpbgfillcolor} (10.638,3.360)--(10.691,3.360)--(10.691,3.787)--(10.638,3.787)--cycle;
\draw[gp path] (10.638,3.360)--(10.638,3.787)--(10.691,3.787)--(10.691,3.360)--cycle;
\gpfill{color=gpbgfillcolor} (10.638,2.825)--(10.691,2.825)--(10.691,3.253)--(10.638,3.253)--cycle;
\draw[gp path] (10.638,2.825)--(10.638,3.253)--(10.691,3.253)--(10.691,2.825)--cycle;
\gpfill{color=gpbgfillcolor} (10.638,2.291)--(10.691,2.291)--(10.691,2.718)--(10.638,2.718)--cycle;
\draw[gp path] (10.638,2.291)--(10.638,2.718)--(10.691,2.718)--(10.691,2.291)--cycle;
\gpfill{color=gpbgfillcolor} (10.638,1.757)--(10.691,1.757)--(10.691,2.184)--(10.638,2.184)--cycle;
\draw[gp path] (10.638,1.757)--(10.638,2.184)--(10.691,2.184)--(10.691,1.757)--cycle;
\node[gp node center] at (2.953,1.704) {$s_t$};
\node[gp node center] at (4.823,1.169) {$s_o$};
\node[gp node center] at (9.276,1.169) {$s_i$};
\node[gp node center] at (4.400,2.98) {$T$};
\node[gp node center] at (9.810,3.840) {$F$};
\node[gp node center] at (5.870,3.980) {$F$};
\node[gp node center, rotate=90] at (1.395,5.844) {Background};
\node[gp node center, rotate=90] at (3.800,5.844) {Refractive plane};
\node[gp node center] at (7.940,7.313) {Lens};
\node[gp node center, rotate=-90] at (10.925,2.905) {Image sensor};
\node[gp node left] at (11.012,5.138) {$\Delta_t$};
\node[gp node left] at (0.959,2.594) {$\delta_b$};
\node[gp node left] at (4.574,5.699) {$\delta_t$};
\gpsetlinetype{gp lt border}
\draw[gp path](1.707,1.704)--(1.707,7.046);
\draw[gp path](4.200,1.704)--(4.200,7.046);
\draw[gp path](7.940,1.704)--(7.940,3.039);
\draw[gp path](7.762,3.039)--(8.118,3.039);
\draw[gp path](7.940,7.046)--(7.940,5.710);
\draw[gp path](7.762,5.710)--(8.118,5.710);
\draw[gp path](10.611,1.704)--(10.611,7.046);
\gpcolor{gp lt color 4}
\draw[gp path,<->](1.707,1.436)--(4.200,1.436);
\draw[gp path,<->](1.707,0.902)--(7.940,0.902);
\draw[gp path,<->](7.940,0.902)--(10.611,0.902);
\draw[gp path,<->](10.878,5.519)--(10.878,4.756);
\draw[gp path,<->](1.520,1.704)--(1.520,3.484);
\draw[gp path,->](4.387,5.699)--(4.387,4.441);
\draw[gp path,->](4.387,1.928)--(4.387,2.278);
\gpcolor{gp lt color 0}
\gpsetlinetype{gp lt plot 1}
\draw[gp path](1.707,2.594)--(7.940,5.710);
\draw[gp path](1.707,2.594)--(7.940,3.039);
\draw[gp path](7.940,5.710)--(10.611,5.138);
\draw[gp path](7.940,3.039)--(10.611,5.138);
\draw[gp path](1.707,2.594)--(10.611,5.138);
\gpcolor{gp lt color 1}
\gpsetlinetype{gp lt plot 3}
\draw[gp path,dotted] (0.460,4.375)--(0.576,4.375)--(0.692,4.375)--(0.808,4.375)--(0.924,4.375)%
  --(1.040,4.375)--(1.156,4.375)--(1.272,4.375)--(1.388,4.375)--(1.504,4.375)--(1.620,4.375)%
  --(1.736,4.375)--(1.852,4.375)--(1.968,4.375)--(2.084,4.375)--(2.200,4.375)--(2.316,4.375)%
  --(2.433,4.375)--(2.549,4.375)--(2.665,4.375)--(2.781,4.375)--(2.897,4.375)--(3.013,4.375)%
  --(3.129,4.375)--(3.245,4.375)--(3.361,4.375)--(3.477,4.375)--(3.593,4.375)--(3.709,4.375)%
  --(3.825,4.375)--(3.941,4.375)--(4.057,4.375)--(4.173,4.375)--(4.289,4.375)--(4.405,4.375)%
  --(4.521,4.375)--(4.637,4.375)--(4.753,4.375)--(4.869,4.375)--(4.985,4.375)--(5.101,4.375)%
  --(5.217,4.375)--(5.333,4.375)--(5.449,4.375)--(5.565,4.375)--(5.681,4.375)--(5.797,4.375)%
  --(5.913,4.375)--(6.029,4.375)--(6.145,4.375)--(6.262,4.375)--(6.378,4.375)--(6.494,4.375)%
  --(6.610,4.375)--(6.726,4.375)--(6.842,4.375)--(6.958,4.375)--(7.074,4.375)--(7.190,4.375)%
  --(7.306,4.375)--(7.422,4.375)--(7.538,4.375)--(7.654,4.375)--(7.770,4.375)--(7.886,4.375)%
  --(8.002,4.375)--(8.118,4.375)--(8.234,4.375)--(8.350,4.375)--(8.466,4.375)--(8.582,4.375)%
  --(8.698,4.375)--(8.814,4.375)--(8.930,4.375)--(9.046,4.375)--(9.162,4.375)--(9.278,4.375)%
  --(9.394,4.375)--(9.510,4.375)--(9.626,4.375)--(9.742,4.375)--(9.858,4.375)--(9.974,4.375)%
  --(10.091,4.375)--(10.207,4.375)--(10.323,4.375)--(10.439,4.375)--(10.555,4.375)--(10.671,4.375)%
  --(10.787,4.375)--(10.903,4.375)--(11.019,4.375)--(11.135,4.375)--(11.251,4.375)--(11.367,4.375)%
  --(11.483,4.375)--(11.599,4.375)--(11.715,4.375)--(11.831,4.375)--(11.947,4.375);
\gpfill{color=gp lt color 0,opacity=0.10} (1.736,3.495)--(1.736,1.710)--(1.851,1.735)--(1.851,3.536)--cycle;
\gpfill{color=gp lt color 0,opacity=0.10} (1.852,3.536)--(1.852,1.735)--(1.967,1.760)--(1.967,3.578)--cycle;
\gpfill{color=gp lt color 0,opacity=0.10} (1.968,3.578)--(1.968,1.760)--(2.083,1.784)--(2.083,3.619)--cycle;
\gpfill{color=gp lt color 0,opacity=0.10} (2.084,3.619)--(2.084,1.784)--(2.199,1.809)--(2.199,3.660)--cycle;
\gpfill{color=gp lt color 0,opacity=0.10} (2.200,3.660)--(2.200,1.809)--(2.315,1.834)--(2.315,3.702)--cycle;
\gpfill{color=gp lt color 0,opacity=0.10} (2.316,3.702)--(2.316,1.834)--(2.432,1.859)--(2.432,3.743)--cycle;
\gpfill{color=gp lt color 0,opacity=0.10} (2.433,3.743)--(2.433,1.859)--(2.548,1.884)--(2.548,3.785)--cycle;
\gpfill{color=gp lt color 0,opacity=0.10} (2.549,3.785)--(2.549,1.884)--(2.664,1.909)--(2.664,3.826)--cycle;
\gpfill{color=gp lt color 0,opacity=0.10} (2.665,3.826)--(2.665,1.909)--(2.780,1.934)--(2.780,3.868)--cycle;
\gpfill{color=gp lt color 0,opacity=0.10} (2.781,3.868)--(2.781,1.934)--(2.896,1.958)--(2.896,3.909)--cycle;
\gpfill{color=gp lt color 0,opacity=0.10} (2.897,3.909)--(2.897,1.958)--(3.012,1.983)--(3.012,3.951)--cycle;
\gpfill{color=gp lt color 0,opacity=0.10} (3.013,3.951)--(3.013,1.983)--(3.128,2.008)--(3.128,3.992)--cycle;
\gpfill{color=gp lt color 0,opacity=0.10} (3.129,3.992)--(3.129,2.008)--(3.244,2.033)--(3.244,4.033)--cycle;
\gpfill{color=gp lt color 0,opacity=0.10} (3.245,4.033)--(3.245,2.033)--(3.360,2.058)--(3.360,4.075)--cycle;
\gpfill{color=gp lt color 0,opacity=0.10} (3.361,4.075)--(3.361,2.058)--(3.476,2.083)--(3.476,4.116)--cycle;
\gpfill{color=gp lt color 0,opacity=0.10} (3.477,4.116)--(3.477,2.083)--(3.592,2.108)--(3.592,4.158)--cycle;
\gpfill{color=gp lt color 0,opacity=0.10} (3.593,4.158)--(3.593,2.108)--(3.708,2.132)--(3.708,4.199)--cycle;
\gpfill{color=gp lt color 0,opacity=0.10} (3.709,4.199)--(3.709,2.132)--(3.824,2.157)--(3.824,4.241)--cycle;
\gpfill{color=gp lt color 0,opacity=0.10} (3.825,4.241)--(3.825,2.157)--(3.940,2.182)--(3.940,4.282)--cycle;
\gpfill{color=gp lt color 0,opacity=0.10} (3.941,4.282)--(3.941,2.182)--(4.056,2.207)--(4.056,4.323)--cycle;
\gpfill{color=gp lt color 0,opacity=0.10} (4.057,4.323)--(4.057,2.207)--(4.172,2.232)--(4.172,4.365)--cycle;
\gpfill{color=gp lt color 0,opacity=0.10} (4.173,4.365)--(4.173,2.232)--(4.288,2.257)--(4.288,4.406)--cycle;
\gpfill{color=gp lt color 0,opacity=0.10} (4.289,4.406)--(4.289,2.257)--(4.404,2.282)--(4.404,4.448)--cycle;
\gpfill{color=gp lt color 0,opacity=0.10} (4.405,4.448)--(4.405,2.282)--(4.520,2.306)--(4.520,4.489)--cycle;
\gpfill{color=gp lt color 0,opacity=0.10} (4.521,4.489)--(4.521,2.306)--(4.636,2.331)--(4.636,4.531)--cycle;
\gpfill{color=gp lt color 0,opacity=0.10} (4.637,4.531)--(4.637,2.331)--(4.752,2.356)--(4.752,4.572)--cycle;
\gpfill{color=gp lt color 0,opacity=0.10} (4.753,4.572)--(4.753,2.356)--(4.868,2.381)--(4.868,4.613)--cycle;
\gpfill{color=gp lt color 0,opacity=0.10} (4.869,4.613)--(4.869,2.381)--(4.984,2.406)--(4.984,4.655)--cycle;
\gpfill{color=gp lt color 0,opacity=0.10} (4.985,4.655)--(4.985,2.406)--(5.100,2.431)--(5.100,4.696)--cycle;
\gpfill{color=gp lt color 0,opacity=0.10} (5.101,4.696)--(5.101,2.431)--(5.216,2.456)--(5.216,4.738)--cycle;
\gpfill{color=gp lt color 0,opacity=0.10} (5.217,4.738)--(5.217,2.456)--(5.332,2.481)--(5.332,4.779)--cycle;
\gpfill{color=gp lt color 0,opacity=0.10} (5.333,4.779)--(5.333,2.481)--(5.448,2.505)--(5.448,4.821)--cycle;
\gpfill{color=gp lt color 0,opacity=0.10} (5.449,4.821)--(5.449,2.505)--(5.564,2.530)--(5.564,4.862)--cycle;
\gpfill{color=gp lt color 0,opacity=0.10} (5.565,4.862)--(5.565,2.530)--(5.680,2.555)--(5.680,4.903)--cycle;
\gpfill{color=gp lt color 0,opacity=0.10} (5.681,4.903)--(5.681,2.555)--(5.796,2.580)--(5.796,4.945)--cycle;
\gpfill{color=gp lt color 0,opacity=0.10} (5.797,4.945)--(5.797,2.580)--(5.912,2.605)--(5.912,4.986)--cycle;
\gpfill{color=gp lt color 0,opacity=0.10} (5.913,4.986)--(5.913,2.605)--(6.028,2.630)--(6.028,5.028)--cycle;
\gpfill{color=gp lt color 0,opacity=0.10} (6.029,5.028)--(6.029,2.630)--(6.144,2.655)--(6.144,5.069)--cycle;
\gpfill{color=gp lt color 0,opacity=0.10} (6.145,5.069)--(6.145,2.655)--(6.261,2.679)--(6.261,5.111)--cycle;
\gpfill{color=gp lt color 0,opacity=0.10} (6.262,5.111)--(6.262,2.679)--(6.377,2.704)--(6.377,5.152)--cycle;
\gpfill{color=gp lt color 0,opacity=0.10} (6.378,5.152)--(6.378,2.704)--(6.493,2.729)--(6.493,5.194)--cycle;
\gpfill{color=gp lt color 0,opacity=0.10} (6.494,5.194)--(6.494,2.729)--(6.609,2.754)--(6.609,5.235)--cycle;
\gpfill{color=gp lt color 0,opacity=0.10} (6.610,5.235)--(6.610,2.754)--(6.725,2.779)--(6.725,5.276)--cycle;
\gpfill{color=gp lt color 0,opacity=0.10} (6.726,5.276)--(6.726,2.779)--(6.841,2.804)--(6.841,5.318)--cycle;
\gpfill{color=gp lt color 0,opacity=0.10} (6.842,5.318)--(6.842,2.804)--(6.957,2.829)--(6.957,5.359)--cycle;
\gpfill{color=gp lt color 0,opacity=0.10} (6.958,5.359)--(6.958,2.829)--(7.073,2.853)--(7.073,5.401)--cycle;
\gpfill{color=gp lt color 0,opacity=0.10} (7.074,5.401)--(7.074,2.853)--(7.189,2.878)--(7.189,5.442)--cycle;
\gpfill{color=gp lt color 0,opacity=0.10} (7.190,5.442)--(7.190,2.878)--(7.305,2.903)--(7.305,5.484)--cycle;
\gpfill{color=gp lt color 0,opacity=0.10} (7.306,5.484)--(7.306,2.903)--(7.421,2.928)--(7.421,5.525)--cycle;
\gpfill{color=gp lt color 0,opacity=0.10} (7.422,5.525)--(7.422,2.928)--(7.537,2.953)--(7.537,5.566)--cycle;
\gpfill{color=gp lt color 0,opacity=0.10} (7.538,5.566)--(7.538,2.953)--(7.653,2.978)--(7.653,5.608)--cycle;
\gpfill{color=gp lt color 0,opacity=0.10} (7.654,5.608)--(7.654,2.978)--(7.769,3.003)--(7.769,5.649)--cycle;
\gpfill{color=gp lt color 0,opacity=0.10} (7.770,5.649)--(7.770,3.003)--(7.885,3.027)--(7.885,5.691)--cycle;
\gpfill{color=gp lt color 0,opacity=0.10} (7.886,5.691)--(7.886,3.027)--(8.001,3.079)--(8.001,5.706)--cycle;
\gpfill{color=gp lt color 0,opacity=0.10} (8.002,5.706)--(8.002,3.079)--(8.117,3.153)--(8.117,5.697)--cycle;
\gpfill{color=gp lt color 0,opacity=0.10} (8.118,5.697)--(8.118,3.153)--(8.233,3.228)--(8.233,5.689)--cycle;
\gpfill{color=gp lt color 0,opacity=0.10} (8.234,5.689)--(8.234,3.228)--(8.349,3.303)--(8.349,5.681)--cycle;
\gpfill{color=gp lt color 0,opacity=0.10} (8.350,5.681)--(8.350,3.303)--(8.465,3.377)--(8.465,5.672)--cycle;
\gpfill{color=gp lt color 0,opacity=0.10} (8.466,5.672)--(8.466,3.377)--(8.581,3.452)--(8.581,5.664)--cycle;
\gpfill{color=gp lt color 0,opacity=0.10} (8.582,5.664)--(8.582,3.452)--(8.697,3.526)--(8.697,5.656)--cycle;
\gpfill{color=gp lt color 0,opacity=0.10} (8.698,5.656)--(8.698,3.526)--(8.813,3.601)--(8.813,5.648)--cycle;
\gpfill{color=gp lt color 0,opacity=0.10} (8.814,5.648)--(8.814,3.601)--(8.929,3.676)--(8.929,5.639)--cycle;
\gpfill{color=gp lt color 0,opacity=0.10} (8.930,5.639)--(8.930,3.676)--(9.045,3.750)--(9.045,5.631)--cycle;
\gpfill{color=gp lt color 0,opacity=0.10} (9.046,5.631)--(9.046,3.750)--(9.161,3.825)--(9.161,5.623)--cycle;
\gpfill{color=gp lt color 0,opacity=0.10} (9.162,5.623)--(9.162,3.825)--(9.277,3.899)--(9.277,5.614)--cycle;
\gpfill{color=gp lt color 0,opacity=0.10} (9.278,5.614)--(9.278,3.899)--(9.393,3.974)--(9.393,5.606)--cycle;
\gpfill{color=gp lt color 0,opacity=0.10} (9.394,5.606)--(9.394,3.974)--(9.509,4.049)--(9.509,5.598)--cycle;
\gpfill{color=gp lt color 0,opacity=0.10} (9.510,5.598)--(9.510,4.049)--(9.625,4.123)--(9.625,5.589)--cycle;
\gpfill{color=gp lt color 0,opacity=0.10} (9.626,5.589)--(9.626,4.123)--(9.741,4.198)--(9.741,5.581)--cycle;
\gpfill{color=gp lt color 0,opacity=0.10} (9.742,5.581)--(9.742,4.198)--(9.857,4.272)--(9.857,5.573)--cycle;
\gpfill{color=gp lt color 0,opacity=0.10} (9.858,5.573)--(9.858,4.272)--(9.973,4.347)--(9.973,5.565)--cycle;
\gpfill{color=gp lt color 0,opacity=0.10} (9.974,5.565)--(9.974,4.347)--(10.090,4.421)--(10.090,5.556)--cycle;
\gpfill{color=gp lt color 0,opacity=0.10} (10.091,5.556)--(10.091,4.421)--(10.206,4.496)--(10.206,5.548)--cycle;
\gpfill{color=gp lt color 0,opacity=0.10} (10.207,5.548)--(10.207,4.496)--(10.322,4.571)--(10.322,5.540)--cycle;
\gpfill{color=gp lt color 0,opacity=0.10} (10.323,5.540)--(10.323,4.571)--(10.438,4.645)--(10.438,5.531)--cycle;
\gpfill{color=gp lt color 0,opacity=0.10} (10.439,5.531)--(10.439,4.645)--(10.554,4.720)--(10.554,5.523)--cycle;
\gpdefrectangularnode{gp plot 1}{\pgfpoint{0.460cm}{0.368cm}}{\pgfpoint{11.947cm}{8.381cm}}
\node[gp node right] at (2.259,2.294) {$B$};
\gpcolor{gp lt color 4}
\draw[thin, <->](7.94,5.7)--(7.94,3.04);
\node[gp node center] at (8.3,5.1) {$A$};
\node[gp node center] at (11.193,6.822) {$\} \,\,\, \ell_{px}$};
\end{tikzpicture}